\documentclass[prd,aps,floats,epsfig,twocolumn]{revtex4}  
 \usepackage{graphicx}
 \usepackage{amssymb}
 \usepackage{ulem}
 \usepackage{multirow}
  \usepackage{color}
 \def\be{\begin{equation}}
\def\ee{\end{equation}}
 \def\bi{\begin{itemize}}
 \def\ei{\end{itemize}}
  \def\ben{\begin{enumerate}}
\def\een{\end{enumerate}}
  \def\bt{\begin{tabular}}
\def\et{\end{tabular}}
\def\bc{\begin{center}}
\def\ec{\end{center}}

\def\hub{{\cal H}}

\def\bea{\begin{eqnarray}}
\def\eea{\end{eqnarray}}

\begin{document}

\input{epsf}

\title{Current constraints on the cosmic growth history}
\author {Rachel Bean}
\author {Matipon Tangmatitham}
\affiliation{Department of Astronomy, Cornell University, Ithaca, NY 14853, USA.}

\begin{abstract}
We present constraints on the cosmic growth history with recent cosmological data, allowing for deviations from $\Lambda$CDM as might arise if cosmic acceleration is due to modifications to GR or inhomogeneous dark energy.  We combine measures of the cosmic expansion history, from Type 1a supernovae, baryon acoustic oscillations and the CMB, with constraints on the growth of structure from recent galaxy, CMB and weak lensing surveys along with ISW-galaxy cross-correlations. Deviations from $\Lambda$CDM are parameterized by phenomenological modifications to the Poisson equation and the relationship between the two Newtonian potentials. We find modifications that are present at the time the CMB is formed are tightly constrained through their impact on the well-measured CMB acoustic peaks. By contrast, constraints on late-time modifications to the growth history, as might arise if modifications are related to the onset of cosmic acceleration, are far weaker, but remain consistent with $\Lambda$CDM at the 95\% confidence level. For these late-time modifications  we find that differences in the evolution on large and small scales could provide an interesting signature by which to search for modified growth histories with future wide angular coverage, large scale structure surveys.
\end{abstract}
\maketitle

%%%%%%%%%%%%%%%%%%%%%%%%%%%%%%%%%%%%%%%%%%%%%%
\section{Introduction}
\label{intro}
%%%%%%%%%%%%%%%%%%%%%%%%%%%%%%%%%%%%%%%%%%%%%%

A central challenge facing cosmology today is to ascertain the origin of current accelerated cosmic expansion (see for example \cite{Copeland:2006wr,Silvestri:2009hh,Caldwell:2009ix}). Precision measurements of geometric distances to astrophysical objects -- the luminosity distance to Type 1a supernovae (SN1a), the ratio of angular diameter and radial distances from baryon acoustic oscillations  in the 3D galaxy distribution (BAO) and the angular diameter distance to last scattering of the Cosmic Microwave Background (CMB) -- have provided consistent evidence that the universe underwent a transition to acceleration rather recently, when the universe was roughly half the size it is today. 

Distance measurements, however, purely probe the homogeneous expansion history. This is only part of the story if  acceleration derives from a modification to General Relativity (GR) on large scales, or a new type of matter, rather than a cosmological constant ($\Lambda$). If this diverse range of possible causes for acceleration are considered, a clearer discrimination between them is gained from also considering  the growth of density fluctuations, and the large scale structures they seed.

Recently there has been significant interest in the potential of upcoming surveys to constrain the growth of structure including multi-frequency imaging, galaxy and weak lensing measurements \cite{Ishak:2005zs,Knox:2006fh,Amendola:2007rr,Schmidt:2007vj,Zhang:2007nk,Jain:2007yk, Zhang:2008ba,Zhao:2008bn,Song:2008vm,Song:2008xd,Guzik:2009cm,Rapetti:2009ri,Zhao:2009fn,Calabrese:2009tt,Schrabback:2009ba},  spectroscopic surveys to measure the peculiar velocity distribution \cite{Wang:2007ht,Song:2008qt,Percival:2008sh,White:2008jy,Song:2008xd,Guzik:2009cm}, and 21cm intensity surveys \cite{Masui:2009cj}. A principle focus is the application auto- and cross-correlations of galaxy, weak lensing and velocity fields to ascertain if signatures of deviations from a standard scenario, of GR and a homogeneous dark energy component, can be detected in the growth history. 

Existing measurements of the growth of structure can already provide constraints on deviations from this standard scenario, however. Studies have included considering phenomenological models of modified gravity \cite{Sealfon:2004gz,Caldwell:2007cw,Wang:2007fsa,Nesseris:2007pa,Dore:2007jh,Daniel:2009kr,Giannantonio:2009gi, Bean:2009wj,Daniel:2010ky}, the presence of dark energy inhomogeneities \cite{Bean:2003fb,Weller:2003hw,Hu:2004yd,Hannestad:2005ak, Corasaniti:2005pq,Koivisto:2005mm,Sapone:2009mb,dePutter:2010vy} and specific modified gravity theories such as the DGP \cite{Dvali:2000hr} scenario  \cite{Sawicki:2005cc, Movahed:2007ie, Guo:2006ce,Xia:2009gb,Fang:2008kc,Lombriser:2009xg} and $f(R)$ gravity \cite{Zhang:2005vt,Amarzguioui:2005zq,Bean:2006up,Song:2007da,Carvalho:2008am,Schmidt:2009am}. 

In this paper we provide a current reference point, of how well existing cosmological data can constrain the growth of structure, against which the future survey targets can be framed. We use a combination of both the current cosmological distance measures and auto- and cross- correlations of large scale structure measurements  to place constraints on the growth history when it is not required to be dictated by GR.  Type 1a supernovae from the Supernovae Cosmology Project  \cite{Kowalski:2008ez} and the CfA3 sample \cite{Hicken:2009dk}, the joint analysis of BAO acoustic signatures in the Two-Degree Field (2dF) and DR7 Sloan Digital Sky (SDSS) surveys \cite{Percival:2009xn}, and the WMAP 7-year (WMAP7) CMB temperature and polarization spectra \cite{Larson:2010gs} provide constraints on the cosmic expansion history. The WMAP 7-year CMB data, in combination with  the matter power spectrum from the SDSS DR7 release \cite{Reid:2009xm}, Integated Sachs Wolfe (ISW)-galaxy cross-correlations with 2MASS and SDSS Luminous Red Galaxy (LRG) galaxy surveys \cite{Ho:2008bz} and the COSMOS weak lensing survey \cite{Massey:2007gh}  constrain the growth history.

We employ a phenomenological framework to describe how the growth history might be modified, rather than choosing to investigate constraints on specific modified gravity models currently proposed in the literature. A constant or simply evolving equation of state is used to investigate observational signatures of the origin of cosmic acceleration in the background expansion history. Here we use simply time- and scale-varying models to bridge the link between observation and theory for the growth history. 

In section \ref{growth} we discuss phenomenological changes to the standard growth history, modifying the perturbed Einstein equations through the Poisson's equation and the relationship between the two Newtonian potentials. 
We also consider four model-independent parameters commonly used to describe cosmic growth history. In section \ref{data} we discuss the data sets used in the analysis and in section \ref{effect} summarize the effects that the modified growth history can have on the observed large scale structure correlations.  In section \ref{mcmc}, we summarize the results of Monte Carlo Markov Chain (MCMC) analyses to place constraints on the cosmic growth history, and draw together our findings and discuss the implications for future work in section \ref{discussion}.

%%%%%%%%%%%%%%%%%%%%%%%%%%%%%%%%%%%%%%%%%%%%%%
\section{Modeling the growth of structure}
\label{growth}
%%%%%%%%%%%%%%%%%%%%%%%%%%%%%%%%%%%%%%%%%%%%%%
We consider the conformal Newtonian gauge to describe the metric
\bea
ds^2 &=&-a(\tau)^2[1+2\psi({\bf x},t)]d\tau^2+ a(\tau)^2[1-2\phi({\bf x},t)]d{\bf x}^2 \hspace{0.5cm}
\eea
where $a$ is the scale factor ($a=1$ today), $\tau$ is conformal time, ${\bf x}$ are comoving coordinates, and $\psi$ and $\phi$ are the two Newtonian potentials describing temporal and spatial perturbations to the metric respectively.  We assume units in which the speed of light $c=1$. 
The spatial element can be broken down into a radial, line of sight, component, written in terms of the comoving distance, $\chi = \int_a^1 d\tau$, and a 2D angular element $d\Omega^2$, $d{\bf x}^2 = \left(d\chi^2+r(\chi)^2d\Omega^2\right)$,
where $r(\chi)$ is the  comoving angular diameter distance.

%%%%%%%%%%%%%%%%%
\subsection{Growth in General Relativity}
\label{GRgrowth}
%%%%%%%%%%%%%%%%%

The growth of inhomogeneities is characterized by the evolution of the fractional over-density distribution, $\delta({k},a)\equiv\rho({ k},a)/\bar{\rho}(a)-1$,  and the divergence of the peculiar velocity, $\theta(k,a)$, where ${ k}$ is the comoving wavenumber. 

We employ energy-momentum conservation in the perturbed fluid to obtain the fluid equations for each type of matter, with the density and peculiar velocities  evolving according to \cite{Ma:1995ey},
\bea
\dot\delta &=& -(1+w)(\theta-3\dot\phi)-3\hub(c_s^2-w)\delta\label{fluid1}
\\
\frac{\dot\theta}{k^2} &=& -\hub(1+3w)\frac{\theta}{k^2}-\frac{\dot{w}}{1+w}\frac{\theta}{k^2} + \frac{c_s^2}{(1+w)}\delta - \sigma+\psi. \ \ \ \  \label{fluid2}
\eea
Here $w=P/\rho$ is  the matter's equation of state and $c_s^2=\partial P/\partial \rho$ the sound speed.

To wholly describe the growth history of density and velocity perturbations (up to initial conditions) two further equations are required. We use the Poisson equation, combining the time-time and time-space components of the perturbed Einstein equations, and the anisotropic space-space component,
\bea
k^2\phi &=& - 4\pi Ga^2\sum\rho_i\Delta_i \label{GRPoisson}
\\
\psi - \phi &=&-   12\pi G  a^2 \sum_i \rho_i(1+w_i)\frac{\sigma_i}{k^2} . \label{GRphipsi} 
\eea
where $\Delta_i \equiv \delta_i + 3\hub(1+w)\theta_i/k^2$ is the rest-frame density perturbation of matter species $i$,  $\hub(a)=\dot{a}/{a}$ is the Hubble expansion factor, where dots represent derivatives with respect to conformal time and $\sigma_i$ the anisotropic shear stress. Equation \ref{GRphipsi} shows that the two potentials become effectively equal  in the matter and dark energy dominated eras, when there are negligible  anisotropic shear stresses.

%%%%%%%%%%%%%%%%%
\subsection{Deviations from the standard growth scenario}
\label{deviations}
%%%%%%%%%%%%%%%%%

We parameterize deviations to the growth history by modifying equations (\ref{GRPoisson}) and (\ref{GRphipsi}) with  two scale and time-dependent functions, $Q(k,a)$ and $R(k,a)$, 
\bea
k^2\phi &=&  -4\pi G Q a^2\sum_i\rho_i\Delta_i\label{EE000i}
\\
\psi -R\phi&=&  -   12\pi G Q a^2 \sum_i \rho_i(1+w)\frac{\sigma_i}{k^2} \label{EEij}. \ \ \ \ \
\eea
We assume throughout that matter remains minimally coupled  to gravity so that the fluid equations (\ref{fluid1}) and (\ref{fluid2}) remain unchanged.  

In the context of modified gravity scenarios, equation (\ref{EE000i}) describes a modified Poisson equation in which the gravitational potential responds differently to the presence of matter, while (\ref{EEij}) allows an inequality between the two gravitational potentials, even at late times when anisotropic shear stresses are negligible, with $\psi\approx R\phi$. Note that the time-evolving modifications (\ref{EE000i}) and (\ref{EEij}) are consistent with the predicted evolution of the superhorizon curvature fluctuations, as discussed for example in \cite{Bertschinger:2006aw}, (\ref{EEij}) simply provides the constraint equation by which both $\phi$ and $\psi$ superhorizon solutions can be found.

Appendix \ref{App-camb} summarizes how the modified growth history is incorporated into the publicly available CAMB code \cite{Lewis:1999bs},  used  to predict  the effect of  the model on CMB and galaxy statistics.

The naming conventions for these phenomenological modifications have been somewhat varied in the literature: $Q$ has been defined in other works as $Q$ \cite{Laszlo:2007td},  $f$ \cite{Stab:2006yuk}, $G_{eff}/G$ \cite{Tsujikawa:2007gd,Song:2008vm}, $g$ \cite{Guzik:2009cm} and $\mu$ \cite{Zhao:2008bn,Giannantonio:2009gi,Daniel:2010ky}, while $R$ is defined as $1/(1+\eta)$ \cite{Tsujikawa:2007gd}, $1/\eta$ \cite{Guzik:2009cm,Song:2008vm},  $1/\gamma$ \cite{Zhao:2008bn,Giannantonio:2009gi}, and $1+\varpi$, \cite{Daniel:2008et,Daniel:2009kr,Daniel:2010ky}. We choose $Q$ and $R$  here in order to avoid confusion with the synchronous gauge metric variable,  $\eta$,  discussed in appendix \ref{App-camb} and growth history variables, $g_0$, $f$, and $\gamma$, introduced in section \ref{param}.

Theoretical models of modified gravity can be described using (\ref{EE000i})  and (\ref{EEij}), for example exhibiting time-dependent, but largely scale-independent variations, as with DGP on sub-horizon scales, in the quasi-static regime, \cite{Lue:2004rj,Koyama:2005kd,Hu:2007pj}, or with both time and scale dependence, for example in $f(R)$ theories \cite{Song:2006ej,Hu:2007pj}.  We consider two phenomenological models in our analysis that allow for both time and scale dependence in the modifications.

The first (Model I) parameterizes $Q$ and $R$ with monotonically evolving, time- and space-dependent functions of $a$ and $k$, with
\bea
Q(k,a)-1 &=&  \left[Q_0e^{-k/k_c}+Q_\infty(1-e^{-k/k_c})-1 \right]a^s \nonumber
\\
R(k,a)-1 &=&  \left[R_0e^{-k/k_c}+R_\infty(1-e^{-k/k_c})-1 \right]a^s \hspace{0.25cm}
 \label{QRdef}
\eea
where  $Q_0,R_0$ and $Q_{\infty},R_{\infty}$ are, respectively, the asymptotic values of the modification on superhorizon and subhorizon scales today.  In specific theories, $Q$ and $R$ are typically a monotonically varying function of time and scale. Here, for simplicity, we assume a constant power law index, $s$, to parameterize the time-variation of the modification.  $k_c$ is a comoving transition scale between the large and small scale behavior that allows us to decouple observations on large and small scales to assess which are providing the principal constraints.

We impose two theoretical priors on the model, one that $Q(k,a)>0$, so that matter is attracted into over-densities, and $R(k,a)>-1$ to ensure that null geodesic paths are bent inwards by gravitational potentials.

Modifications to gravity could be mimicked by modifications to the growth of inhomogeneities in the `dark sector'  \cite{Kunz:2006ca,Bertschinger:2008zb}, since the effect of any additional inhomogeneities is only inferred indirectly through their gravitational effect. Such modifications could include the presence of dark energy density, peculiar velocity and anisotropic shear stress fluctuations or non-minimal matter-dark energy interactions that contribute to the right hand side of Einstein's equations. In this context, the inhomogeneities would give rise to effective $Q$ and $R$ parameters,
\bea
Q_{eff} &=&  1+\frac{\rho_{de}\Delta_{de}}{\rho_m\Delta_m}
\\
R_{eff} &=& 1+ \frac{3\rho_{de}}{k^2}\left[\frac{(1+w_{de})\sigma_{de}+  \left( \frac{\Delta_{de}}{\Delta_m}  \right) (1+w_m)\sigma_m}{\rho_m\Delta_m+\rho_{de}\Delta_{de} }\right]\nonumber
\\ &&
\eea
where $\rho_m\Delta_m$ is the net inhomogeneity from Standard Model and cold dark matter, $\rho_{de}\Delta_{de}$ describes the additional dark energy and coupled dark energy-dark matter inhomogeneities, and $\sigma_{de}$ is a dark energy anisotropic shear stress.

If the dark component producing the inhomogeneities is non-relativitic, then it is extremely difficult to sustain a substantial anisotropic shear stress (see for example \cite{Hu:1998kj}). If, in addition, at early times, the dark energy density inhomogeneities are small  in comparison to those of normal matter, then one may reasonably assume $R_{eff}=1$.

The second model (Model II) fixes $R=1$ and modifies the growth history purely through including an extra, additive, time- and space-dependent contribution to the Poisson equation
\bea
k^2\phi &=&-\left(4\pi G a^2\rho_m\Delta_{m} + \frac{3H_0^2}{2}\Delta_{xH}\left(\frac{k}{\hub}\right)^{n_x}a^{s_x}e^{-k/k_x}\right)\hspace{0.75cm} \label{delxdef}
\eea
where $\Delta_{xH}$ parameterizes the amplitude of the modification on horizon scales ($k=\hub^{-1}$) today, $s_x$ and $n_x$  describe the temporal and large scale spectral evolution, and on scales $k>k_x$ the modification is suppressed ($k_x$ would relate to an effective comoving sound horizon, if the modification was due to an additional component). 
As an illustration,  a modification with $n_x=n_s/2$ ($n_s$ is the inflationary spectral index), $s_x=0$ and $k_x=\infty$ would mimic large scale CDM density perturbations in the CDM dominated era. 

In the presence of   pressureless matter this modification is equivalent to
\bea
Q &=& 1+ \frac{\Delta_{xH}\left(\frac{k}{\hub}\right)^{n_x}e^{-k/k_x}}{\Omega_m\Delta_{m}(a) }a^{s_x+1}\label{Qdepert}
\eea
where $\Omega_m$ is the fractional matter density today.

The evolution of an additional  dark component could be described using phenomenological models for macroscopic variables, such as the dark energy equation of state and sound speed \cite{Hu:1998kj}, however this leads to associated theoretical priors on the growth history \cite{Song:2010rm}. 
Here we make no assumptions, apriori, about the origin of the modifications in Models I and II, whether gravitational or from some sort of inhomogeneous matter  \cite{2008arXiv0812.0376C}.

%%%%%%%%%%%%%%%%%%%%%%%%%%%%%%%%%%%%%%%%%%%%%%
\subsection{Parameterizing the growth history}
\label{param}
%%%%%%%%%%%%%%%%%%%%%%%%%%%%%%%%%%%%%%%%%%%%%%

The constraints on the phenomenological model parameters  give a simple indication of whether deviations from GR, described by the modified Poisson equation and inequality between $\phi$ and $\psi$, are allowed by the data. It's instructive, however, to  also consider how these models affect the growth of structure, using parameters that are more model independent. Here we consider four such, gauge-invariant parameters:
\begin{itemize}

\item The total matter power spectrum today
\bea
P(k) =  2\pi^2\left(\frac{h}{k}\right)^3\Delta_{tot}^2
\eea
where $\Delta_{tot}=(\rho_c\Delta_c+\rho_b\Delta_b)/(\rho_c+\rho_b)$ is the gauge invariant,  rest frame, over-density for CDM and baryons combined.

\item The rest-frame CDM over-density growth factor relative to today, $g_0$,
\bea
g_0(k,a) &\equiv & \frac{\Delta_c(k,a)}{\Delta_c(k,a=1)}. 
\eea

\item The growth rate of rest-frame CDM over-densities, parameterized by  the effective power law index $f$, 
\bea
f(k,a) &\equiv& \frac{d\ln \Delta_c(k,a)}{d\ln a}.
\eea
In GR $f=1$  in the CDM era, and $f<1$ at late times in response to accelerated expansion.

\item The change in the growth rate, relative to the change in the matter density
\bea
\gamma(k,a)  \equiv \frac{ \ln f(k,a)}{\ln \Omega_m(a)}.
\eea

$\gamma$ takes a well-constrained value $\approx 0.55$ at late times in $\Lambda CDM$ \cite{Wang:1998gt}. Because of this it, in particular, has been widely discussed as a useful parameter with which to search for deviations from a standard growth history \cite{Linder:2005in,Linder:2007hg,Polarski:2007rr}.
\ei

%%%%%%%%%%%%%%%%%%%%%%%%%%%%%%%%%%%%%%%%%%%%%%
\section{Description of  datasets used}
\label{data}
%%%%%%%%%%%%%%%%%%%%%%%%%%%%%%%%%%%%%%%%%%%%%

To constrain the background expansion history, we use the combined analysis of the `Union' set of supernovae from the Supernovae Cosmology Project \cite{Kowalski:2008ez}, and additional supernovae sample of \cite{Hicken:2009dk}, the joint analysis of BAO acoustic signatures in the Two-Degree Field (2dF) and DR7 Sloan Digital Sky (SDSS) surveys \cite{Percival:2009xn}, and the WMAP 7-year CMB temperature and polarization data \cite{Larson:2010gs}. 

To constrain the growth of structure, we consider the large scale ISW measurements from WMAP7, the matter power spectrum from the SDSS DR7 release \cite{Reid:2009xm}, the 2MASS and SDSS Luminous Red Galaxy (LRG) galaxy auto-correlations and ISW-galaxy cross-correlations selected by Ho. et. al. \cite{Ho:2008bz} and the publicly available data from the COSMOS weak lensing survey \cite{Massey:2007gh}. We provide more detail on how these datasets are employed here.

Under the Limber approximation, the 2D angular power spectrum for the correlation between two fields, $X$ and $Y$, is
\bea
C^{XY}_l &= &\int_0^{\chi_{\infty}} \frac{d\chi}{\chi^2}W_X(\chi)W_Y(\chi)T_X(k_l, \chi)T_Y(k_l, \chi)\Delta_{R}^{2}(k_l), \nonumber \\ && \hspace{0.25cm} 
\eea
where $X,Y$= I,g and $\kappa$ for ISW, galaxy and lensing convergence fields respectively, $W_X$ is the window function associated with the field $X$, and $T_X$ is the transfer function, with $k_l = (l+1/2)/\chi$. 
 
For the ISW and galaxy correlations, we use  data compiled by Ho et al.\cite{Ho:2008bz} for 2MASS and SDSS LRG surveys, with each broken into 4 and 2 redshift bins, 2MASS0-3 and LRG0-1, respectively. These are cross correlated with WMAP5 CMB temperature data.  Following \cite{Ho:2008bz},  we model the non-linear corrections and redshift-dependent bias for each galaxy bin (for the correlations in the Ho et al. samples only) using a Q-model for the matter power spectrum \cite{Cole:2005sx},
\bea
P_{obs}(k) &=& b_{norm}^2b_{rel}(z)^2 \frac{1+Q_gk^2}{1+A_gk}P_{lin}(k),
\label{Qbeq}
\eea
 $b_{rel}(z)$ is a fixed, unit normalized function describing the redshift variation of the galaxy bias in each bin based on their luminosities \cite{Tegmark:2006az}, and $A_g=1.7h^{-1}Mpc$ fits numerical simulations of the galaxy power spectrum \cite{Cole:2005sx}. 

The transfer functions for ISW, galaxy and weak lensing sources are 
\bea
T_{I}&=& e^{-\tau_{reion}}\left(\dot{\tilde\phi}+\dot{\tilde\psi}\right),
\\
T_{g} &=&\sqrt{\frac{1+Q_gk^2}{1+A_gk}}\tilde{\Delta}_c,
\\
T_{\kappa} &=& -\frac{k^2}{2}(\tilde{\phi}+\tilde{\psi}),
\eea
where $\tau_{reion}$ is the reionization optical depth, $\tilde{X}$ is the transfer function of $X$, normalized so that $X^2(k,\chi) = \tilde{X}^2 (k,\chi)\Delta_{\cal R}^2(k)$ where $\Delta_{\cal R}^2(k)$ is the dimensionless primordial spectrum of curvature fluctuations.

The galaxy and lensing window functions are dependent on the  distribution of galaxy number density in each redshift bin $i$ for the relevant survey, $n^{i}(\chi)$,
\bea
W^i_{g}(\chi) &=& b_{rel}^i(\chi)n^i(\chi),
\\
W^i_{\kappa}(\chi) &=& \int_{\chi}^{\chi_{\infty}} d\chi' n^i(\chi') \frac{r(\chi)r(\chi'-\chi)}{r(\chi')},
\eea  
where $n^i$ is normalized such that,
$\int_{0}^{\chi_{\infty}}d\chi n^i(\chi) = 1.$
The bias-weighted mean redshifts $\langle z\rangle=\int dz b_{rel}(z)n(z)$ for the 2MASS0-3 and LRG0-1 samples are  0.06, 0.07, 0.10,  0.12, 0.31 and 0.53 respectively. For ISW, $W_{I}(\chi) = 1$.

Following \cite{Ho:2008bz}, as summarized in Appendix \ref{App-marg}, we treat $b_{norm}$ and $Q_g$ as free parameters that are analytically marginalized  over  \cite{Lewis:2002ah} using the galaxy auto-correlations. This optimal value of the bias is then used to normalize the ISW-galaxy cross-correlation spectrum, $C_{obs}^{Ig}(l) = b_{norm}C^{Ig}(l)$.

We use the publicly available analysis of the COSMOS weak lensing survey data \cite{Massey:2007gh}, given in terms of the projected power spectrum of 2D shear correlations,
\bea
C_{1,2}(\theta)  &\equiv&C_0(\theta) \pm C_4(\theta), 
\\
C_{0,4}(\theta) &\equiv & \frac{1}{2\pi} \int_0^\infty dl  \ l C_l^{\kappa\kappa} J_{0,4}(l\theta).
\eea
The data is broken up into 3 redshift bins, $z<1$, $1<z<1.4$ and $1.4<z<3$. 
 To obtain the weak lensing convergence power spectrum, $C_l^{\kappa\kappa}$, we apply the Smith et al. \cite{Smith:2002dz} non-linear correction to the linear power spectrum.  We use the weak lensing likelihood code written by Julien Lesgourgues \cite{Lesgourgues:2007te} and follow their modeling of the systematic errors in the weak lensing data. Three systematic errors parameters are explicitly marginalized over: $A$ accounts for a 6\% uncertainty in the overall calibration of the shear measurement based on simulated HST images, it has a Gaussian prior centered on 1 with $\sigma_A$ = 0.06; $B$ accounts for a 5\% relative calibration uncertainty between the shear measured in the high and low redshift bins due to underestimation of shear in faint or small galaxies, it has a one-sided Gaussian prior with $\sigma_B=0.05$ and $B\ge 1$; and a 10\% intrusion of low-redshift galaxies into the high redshift bins due to catastrophic photometric redshift errors is modeled by a third parameter, $C$, using a a one-sided Gaussian prior with $\sigma_C=0.10$ and $C\ge 1$.  The correlation functions $C_{1,2}$ are multiplied by $(A/B)^2$, $A^2$ and $(AB)^2/C$ respectively in the low, medium and high redshift bins.

%===========================FIGURE 1 ==================== 
 \begin{figure}[!t]
  \begin{center}
          \includegraphics[width=3.5in,angle=0]{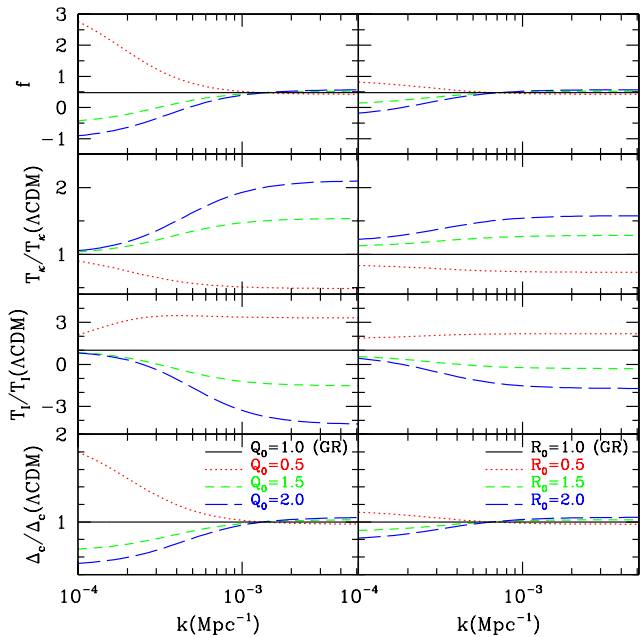}

    \caption{The effect of changing  $Q$ [left panels] and $R$  [right panels]  in Model I, defined in (\ref{QRdef}).  Shown are the CDM density fluctuation, $\Delta_c$, its growth rate, $f$, and ISW and  weak lensing  transfer functions, $T_I$ and $T_{\kappa}$, today, relative to their values for $\Lambda$CDM. Four values of $Q_0$ and $R_0$ are shown for a time-varying but scale-independent modification with $s=3$ and $k_c=\infty$, and all other cosmological parameters fixed. On subhorizon scales, boosting the gravitational potential ($Q>1$) boosts the growth of dark matter density perturbations and the lensing potential. If $Q$ is increased,  on subhorizon scales the boost in the gravitational potential opposes the late-time decay induced by accelerated expansion. The ISW signal can be progressively reduced, canceled out (for $Q$ a little larger than unity) and turn negative as $Q$ increases. The modification has a larger and opposite effect on CDM growth on horizon scales, where the time evolution of the modification plays a key role. Increasing $Q$ causes growth in $\Delta_c$ to slow, or even decay at late times on the largest scales (with $f=d\ln\Delta_c/d\ln a<0$).  The effect of increasing $R$ is quantitatively degenerate for subhorizon CDM growth and qualitatively similar otherwise, but with a slightly reduced amplitude. \label{fig1}} 
  \end{center}
 \end{figure}
  %========================================================

%===========================FIGURE 2 ==================== 
 \begin{figure}[!t]
  \begin{center}
          \includegraphics[width=3.5in,angle=0]{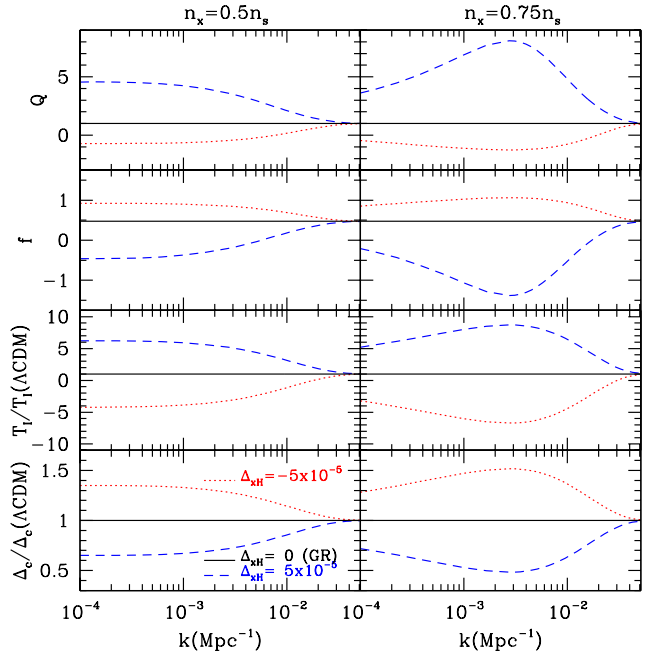}

    \caption{The effect of changing $\Delta_{xH}=0 (GR), \pm 5\times 10^{-5}$ in Model II defined in (\ref{delxdef}) for two different scale dependences, $n_x=0.5n_s$ [left panels] and $n_x=0.75n_s$ [right panels]. For this illustration we fix $s_x=1$  and include a cut off, $k_x=0.01 Mpc^{-1}$, and all other cosmological parameters fixed.  Shown are the CDM density fluctuation, $\Delta_c$, its growth rate, $f$, the  ISW transfer function, $T_I$, and equivalent value of $Q$, today, relative to their values for $\Lambda$CDM.  A positive $\Delta_{xH}$ is equivalent to $Q>1$, suppressing cold dark matter growth, that can give rise to $f<0$ on large scales, and boosting the ISW component. For $n_x=0.5n_s$ the modification has roughly the same spectral evolution as the CDM perturbations, leading to a scale independent effect on large scales. Increasing the spectral tilt, $n_x$, exacerbates the impact of the modification for scales just below the horizon. \label{fig2}} 
  \end{center}
 \end{figure}
  %========================================================

%===========================FIGURE 3 ==================== 
 \begin{figure}[!t]
  \begin{center}
      \includegraphics[width=3.5in,angle=0]{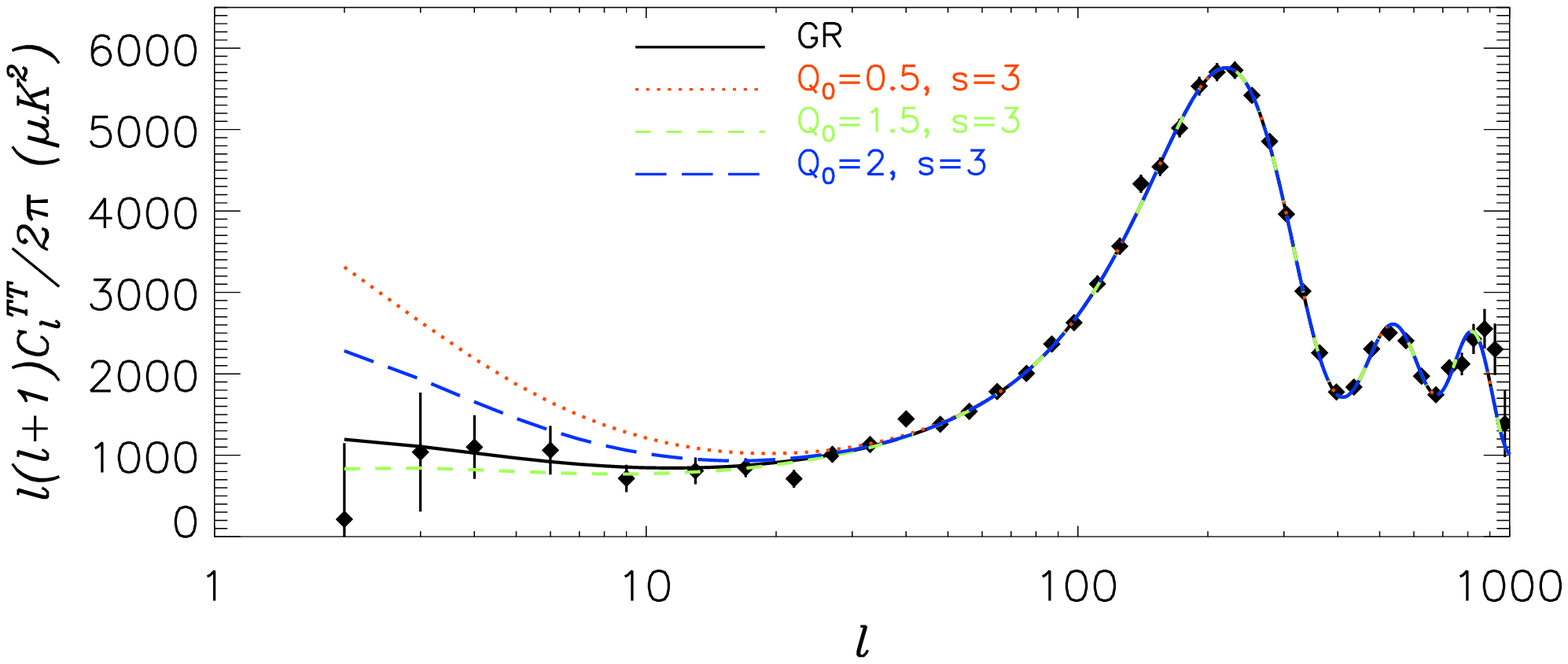}
          \includegraphics[width=3.5in,angle=0]{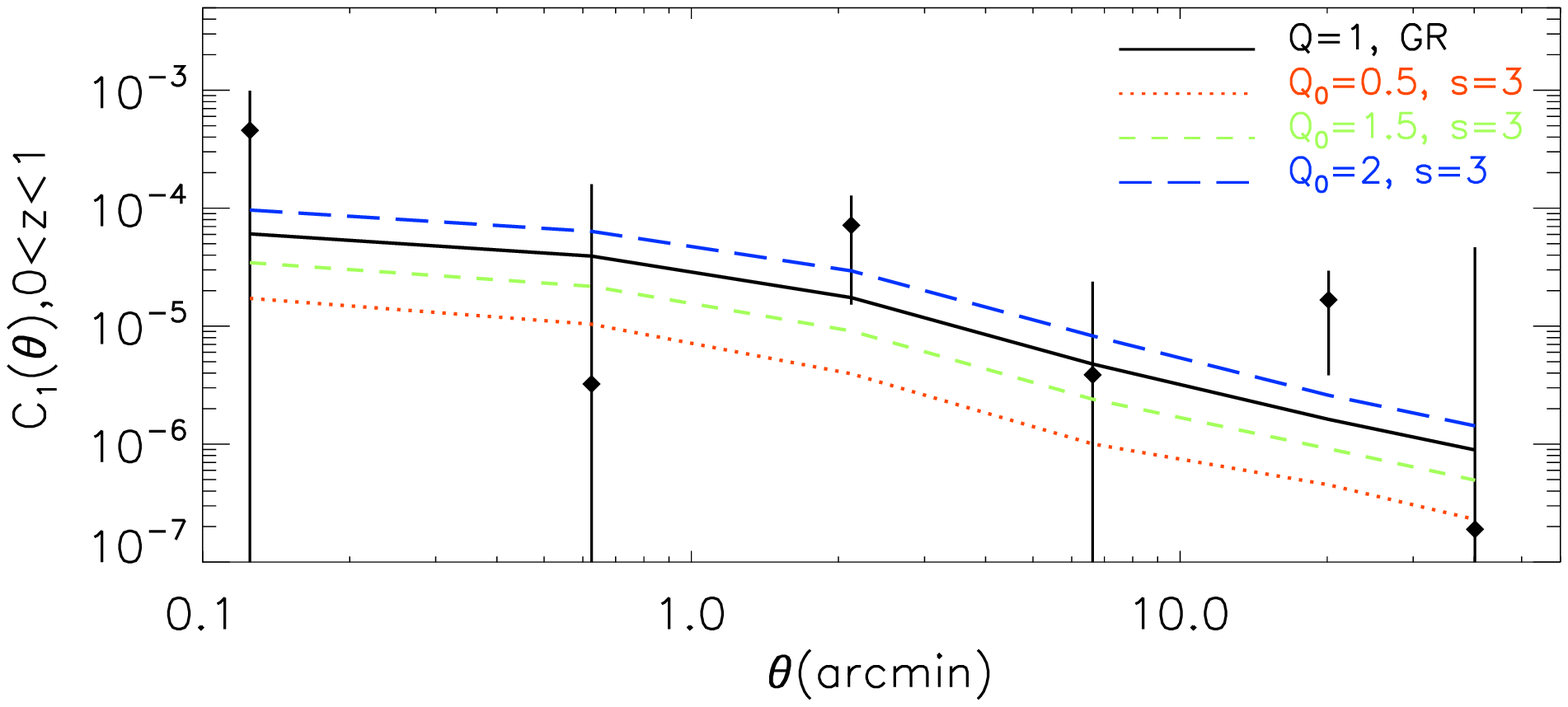}
    \includegraphics[width=3.5in,angle=0]{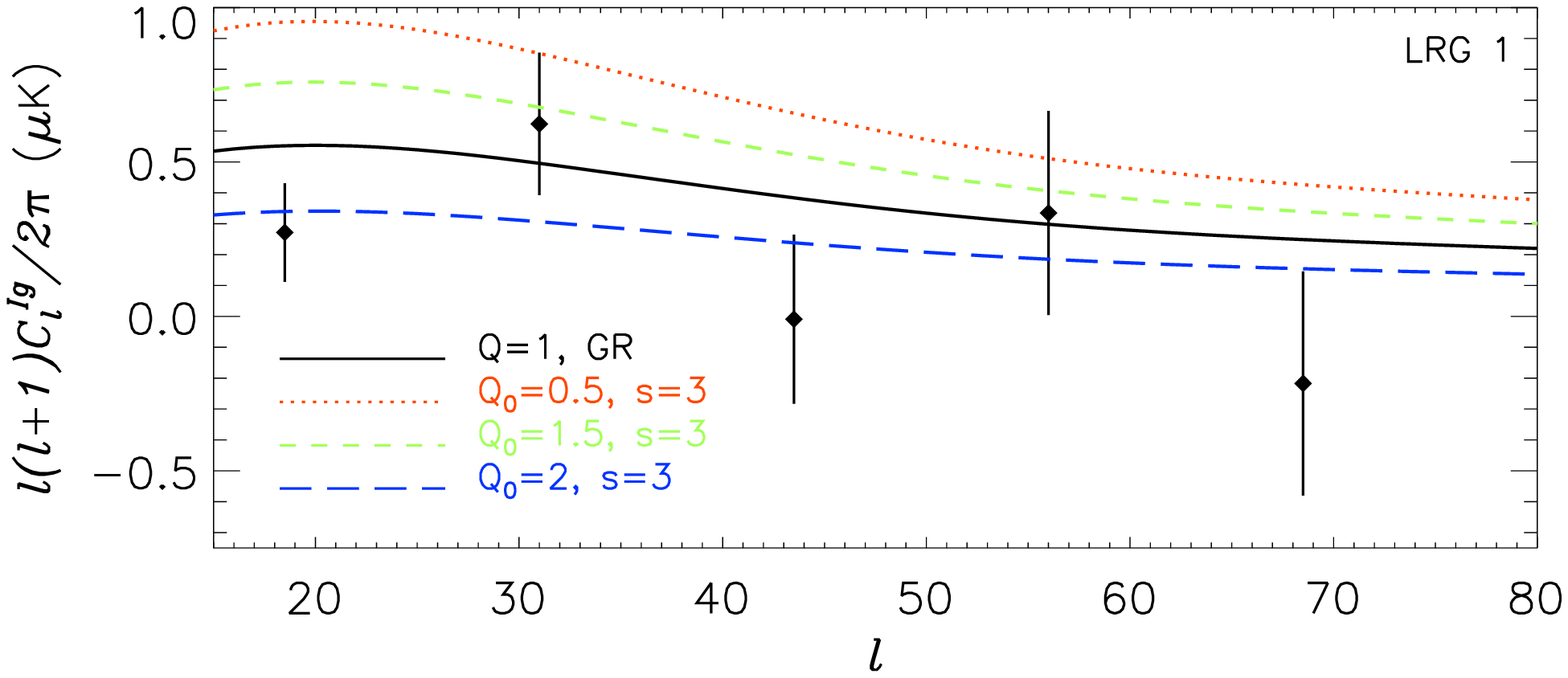}

    \caption{ The effect of the scale-independent and time-evolving modifications in figure \ref{fig1}  on the CMB [upper panel],  lensing correlation [middle panel]  and   ISW-galaxy cross-correlation  [lower panel]. For the lensing and ISW-galaxy correlations example redshift bins are shown. With increasing $Q$,  the lensing potential, and galaxy number density  are boosted while, by contrast, the late-time ISW is suppressed. Both positive ($Q<1$) and negative ($Q>1$)  ISW amplitudes give rise to a boost in the CMB power spectrum at large scales. Increasing $Q$ reduces the ISW-galaxy cross-correlation, and can lead to anti-correlation at low redshifts (this is relevant for the 2MASS samples, but not the higher redshift LRG sample shown here). Increasing $R$ has a qualitatively degenerate effect on the spectra. \label{fig3}}
  \end{center}
 \end{figure}
  %========================================================
  
   %===========================FIGURE 4 ==================== 
 \begin{figure}[!t]
  \begin{center}
          \includegraphics[width=3.5in,angle=0]{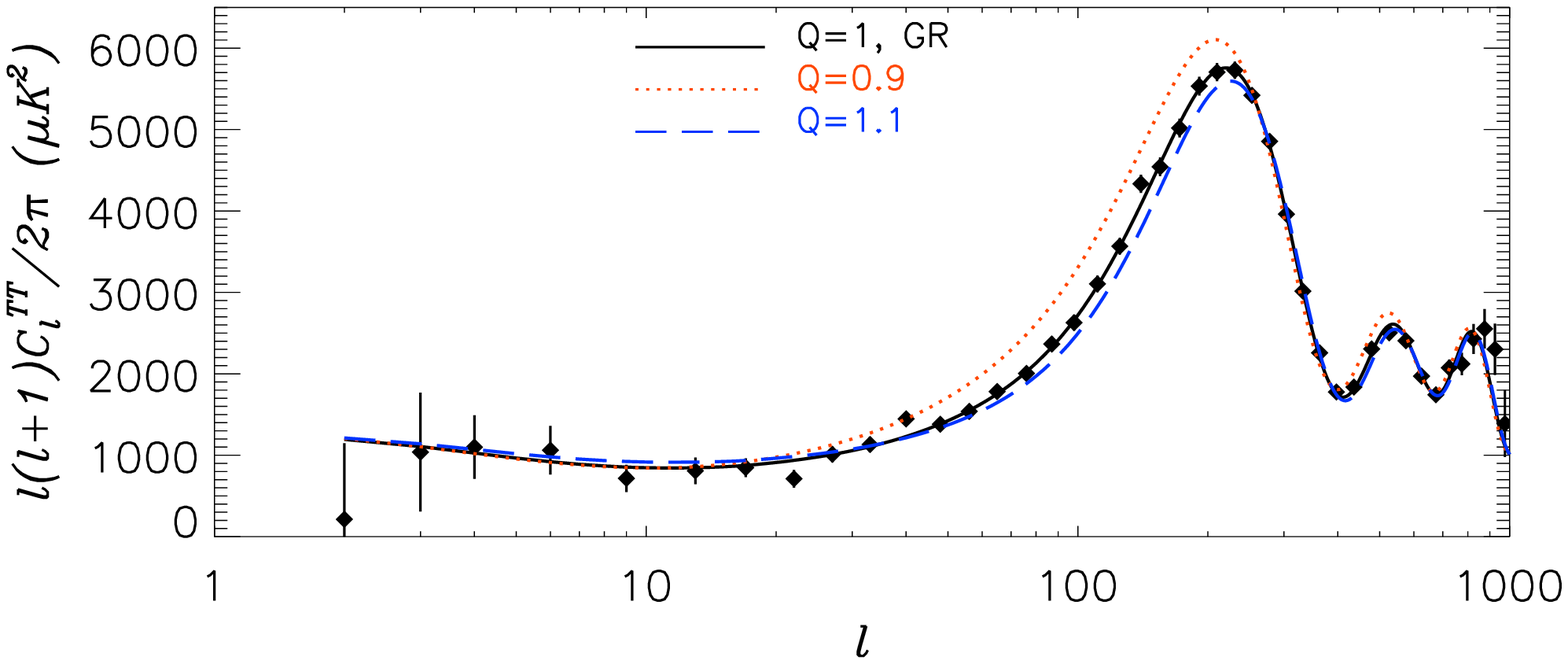}
          \includegraphics[width=3.5in,angle=0]{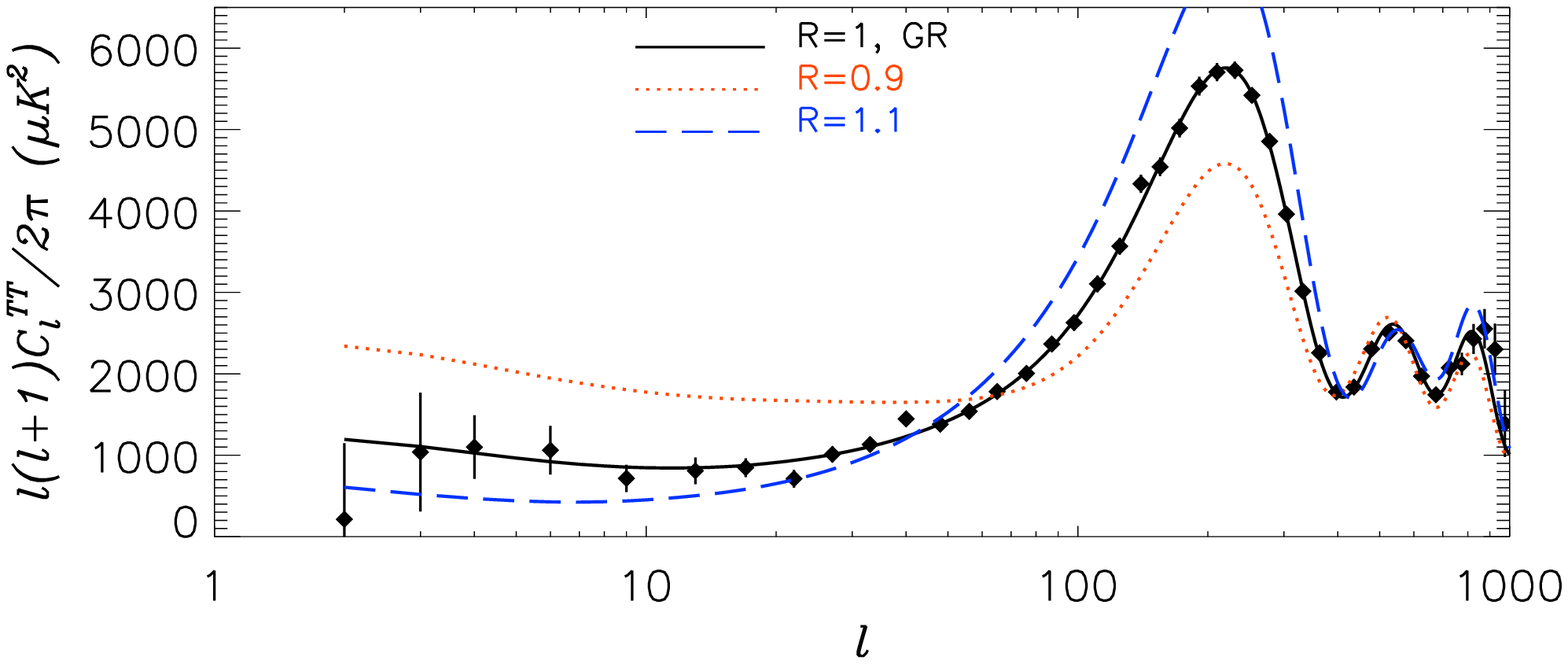}

    \caption{ The effect on the CMB temperature power spectrum of  a time-independent modification to the Poisson equation [top panel] and the relationship between the two potentials, $\phi$ and $\psi$ [lower panel]. The principle effect of a larger $Q$  is to suppress the early ISW effect  on scales comparable with the horizon scale at last scattering. A larger $R$ generates a late-time ISW amplitude even in the matter dominated era, by introducing a multiplicative difference between $\dot\phi$ and $\dot\psi$ and boosts the temperature dipole anisotropy, raising the acoustic peaks. In contrast to the time varying evolution in figure \ref{fig3}, constant $Q$ and $R$ have opposing, rather than degenerate, effects on the CMB.\label{fig4}}
  \end{center}
 \end{figure}
  %========================================================
%%%%%%%%%%%%%%%%%
\section{Observational impact of a modified growth history}
\label{effect}
%%%%%%%%%%%%%%%%%

We can  combine the CDM fluid equations, (\ref{fluid1}) and (\ref{fluid2}) for $w=c_s^2=0$, to give a second order evolution equation for $\Delta_c$ that makes no assumptions about the underlying gravitational theory,
\bea
\ddot\Delta_c&+& \hub\left(1-\frac{\dot{f}_1}{\hub f_1}\right)\dot\Delta_c +f_1\psi= \nonumber 
\\ &&+ 3\left(\ddot\phi+\hub\dot\psi+\hub\left(1-\frac{\dot{f}_1}{\hub f_1}\right)(\dot\phi+\hub\psi)+\dot\hub\psi\right) \hspace{0.75cm} \label{Delcgrow}
\eea
with
\bea
f_1 = k^2 + 12\pi G a^2\sum_i(\rho_i+P_i).
\eea

On sub-horizon scales (\ref{Delcgrow}) reduces to the commonly considered equation for CDM growth,
\bea
\ddot\Delta_c +\hub\dot\Delta_c+k^2\psi & \approx& 0.
\eea
In the absence of anisotropic shear stresses, $k^2\psi \approx -4\pi G a^2 Q R \Delta$. 
  Individually increasing $Q$ or $R$ would have degenerate effects on $\Delta_c$, countering the suppression in growth of CDM over-densities brought on by cosmic acceleration, increasing the lensing shear correlations and reducing the ISW component of the CMB temperature perturbation. As $Q$ or $R$ increases, the correlation between ISW and the galaxy distribution decreases and can give rise to an anti-correlation at late times.

On super-horizon scales, the CDM evolution is determined by the time derivatives of the potentials,
\bea
\ddot\Delta_c+2 \hub\dot\Delta_c- 3\left(\ddot\phi+\hub(2\dot\phi+\dot\psi)+(2\dot\hub+\hub^2)\psi \right) \approx 0.\hspace{0.3cm} 
\eea
On these scales, a modification that boosts the potential can lead to a decline in the growth rate of $\Delta_c$ and the lensing potential, and an associated boosting of the ISW transfer function. 

Figure \ref{fig1} illustrates the effect of the modifications, $Q$ and $R$, on sub- and super-horizon scales for Models I. One finds a marked contrast of the responses of $\Delta_c$ and the transfer functions on large and small scales consistent with the discussion above. On small scales boosting the potential promotes CDM growth, boosting the lensing transfer function, and suppressing the ISW effect. Large scales show the opposite behavior, with CDM growth suppressed by a boost in the potential. Note that on horizon scales suppression of CDM growth can give rise to a negative growth rate, $f<0$, for which the $\gamma$ parameter is undefined. The opposing behaviors on large and small scales suggest that  measuring the growth history on all scales could reveal distinctive signatures of a modified growth history.

Figure \ref{fig2} illustrates how having the additive modification to the Poisson equation, in Model II, modifies the growth history on large scales. Qualitatively the effect is similar to that for Model I, dependent on the sign of $Q$.  $\Delta_{xH}>0$ gives $Q>0$, suppressing CDM growth and boosting the ISW effect. A key difference between Models I and II is the spectral dependence in Model II, parameterized by $n_x$. A spectral index, $n_x\approx n_s/2$, matches the scale behavior of CDM for a scale invariant primordial spectrum and hence just acts to give a scale independent boost to growth on large scales. Introducing a stronger scale dependence accentuates the impact of the modification on sub-horizon scales (for scales above the cut-off scale $k_x$).  This can lead to over-densities being transformed into voids for scales just below the horizon, giving rise to a sharp asymptote and change in sign in $Q$, as $\Delta_c$ transitions from positive to negative. This is distinctly different to the monotonically varying $Q$ in Model I.

  In figure \ref{fig3} we illustrate the effect of a time-evolving modification on the CMB,  lensing and ISW-galaxy correlations. Since the CMB power spectrum is effectively related to the magnitude of the ISW signal, the spectrum does not have a monotonic  response to $Q$ and $R$; there is a minimum ISW contribution for $Q$ or $R$ a little greater than unity, when the suppression in gravitational potential growth rate due to cosmic acceleration is most closely countered by the boosted from the modification, an effect seen in previous modified gravity analyses \cite{Daniel:2008et,Song:2007da}. For larger values of $Q$ or $R$, a negative late-time ISW signal gives rise to a boost in the large scale CMB power spectrum. The lensing power spectrum is similarly not monotonically responsive to $Q$ and $R$,  having a minimum for $R=-1$; however  this degeneracy is removed through our  theoretical prior $R\ge -1$.

The CMB power spectrum alone cannot distinguish between boosting ($Q,R>1$) or suppression ($Q,R<1$) of the gravitational potential since both boost the large scale power. ISW-galaxy correlations vary monotonically with $Q$ and $R$, however, and break  this degeneracy.    Boosting the gravitational potential promotes the formation of large scale structure while suppressing the ISW signal,  hence as $Q$ or $R$ increase there is a monotonic suppression of the cross-correlation that can lead to anti-correlation at low redshift.  

The late-time modification to the growth history  in this analysis leaves a degeneracy between $Q$ and $R$ unbroken (the data is principally sensitive to the sum of the potentials on smaller scales, or its rate of change on larger scales, sensitive to$Q(1+R)/2$). An increase in $Q$ can be countered by a decrease in $R$. In order to break this degeneracy additional correlations would have to be employed, such as using galaxy-peculiar velocity correlations or galaxy-lensing correlations, and considering ratios of observables \cite{Zhang:2007nk,Zhang:2008ba}.

In figure \ref{fig4} we show how time-independent $Q$ or $R$ have opposing, rather than complementary, effects on the first CMB acoustic peak. In the standard GR scenario during the matter dominated era, the gravitational potential is constant and no ISW signal is generated.  An early ISW effect is present around recombination however, because the universe is not purely matter dominated causing the gravitational potential to decay. This decay boosts the CMB power spectrum on scales of order the horizon size at last scattering. A constant $Q>1$ counters the early ISW and suppresses the CMB temperature correlations below a degree scale.  Because it is constant, and parity between $\phi$ and $\psi$ is maintained at late times, it has an insignificant effect on the late time ISW.
Increasing $R$ has two effects: firstly it introduces a constant disparity between the growth rate of the two potentials which leads to a non-zero ISW signal  even during the matter dominated era, and secondly it increases the dipole anisotropy in the CMB, boosting the acoustic peaks. In contrast to the late-time modifications to the growth history,  any increase in a time-independent $Q$ could be countered by an increase in $R$, to minimize the impact on the  well-measured height of the first peak. We willfind that, consistent with this, the data effectively measures the combination $Q-R$.

  %=============================TABLE 1=======================
      \begin{table}[t!]
 \caption{\label{tab1} 1D marginalized 95\% confidence limits for modified gravity parameters $Q$ and $R$, and principle degeneracy direction $Q-R$, for time- and scale-independent deviations from GR in Model I. }

\bt{|c|c|c|c|}
\hline
 Data analyzed & $Q$ & $R$ &$Q-R$
\\ \hline 
All
& [0.97,  1.01]
& [ 0.99,  1.02]
&[-0.032,  0.009] 
\\ 
No WL
& [0.96,  1.01]
& [ 0.98,  1.02]
&[-0.034,  0.008] 
\\ 
No WL, no ISW-galaxy
& [0.98,  1.04]
& [ 0.99,  1.03]
&[-0.025,  0.014] 
\\ \hline
\et
\end{table}
%====================================================

  %=============================TABLE 2=======================
  
    \begin{table*}[t!]
 \caption{\label{tab2} 1D marginalized 95\% confidence limits on the modified gravity parameters $Q_0$, $R_0$ and $Q_\infty$, $R_\infty$, in Model I and principle degeneracy direction, well described by $Q(1+R)/2.$ }

\bt{|c|c|c|c|c|c|c|c|c|}
\hline
$k_c (Mpc^{-1})$  & Data analyzed & $Q_0$ & $R_0$ &$Q_0(1+R_0)/2$& $Q_\infty$ & $R_ \infty $ & $Q_\infty(1+R_\infty)/2$
\\ \hline
\multirow{6}{*}{$\infty$}
&All
&[0.96,  1.60]
&fixed=1
&[0.96,  1.60]
&-
&-
&-
\\ 
&All
&fixed=1
&[0.90,  2.26]
&[0.95,  1.63]
&-
&-
&-
\\ 
&All
&[1.04,  2.66]
&[-0.22,  1.44]
& [0.91,  1.54]
&-
&-
&-
\\ 
 & No WL
&[1.01,  2.77]
&[-0.21, 1.73]
& [0.92, 1.62]
&-
&-
&-
\\ 
&No WL, no ISW-galaxy
& [0.60, 2.55]
&[-0.20, 2.29]
& [0.81, 1.68]
&-
&-
&-
\\ \hline
\multirow{3}{*}{$10^{-2}$}
& All
& [1.06, 3.78]
& [-0.92, 1.91]
& [0.13, 2.00]
&fixed=1
&fixed=1
&-
\\
& All
&fixed=1
&fixed=1
& -
& [1.19,  2.83]
&[-0.47, 1.61]
& [0.68, 1.57]
\\ 
& All
& [0.47,  3.49]
&[-0.80, 2.52]
& [0.28, 1.96]
& [0.97,  2.65]
&[-0.43, 1.76]
& [0.66, 1.58]
\\ \hline
 \et
 \end{table*}
%====================================================

  %%%%%%%%%%%%%%%%%%%%%%%%%%%%%%%%%%
\section{MCMC analysis}
\label{mcmc}
%%%%%%%%%%%%%%%%%
We perform Monte Carlo Markov Chain (MCMC) analyses using a version of the CosmoMC code \cite{Lewis:2002ah} modified to include general growth evolution models described in section \ref{growth}. In scalar tensor theories of gravity, such as $f(R)$,  one can construct the modification to GR that exactly matches a chosen background expansion history \cite{Multamaki:2005zs,Capozziello:2006dj,Nojiri:2006gh,delaCruzDombriz:2006fj}, but gives rise to modifications of the growth of perturbations \cite{Song:2006ej}. With this in mind we assume  a standard $\Lambda$CDM expansion history and focus on the constraints on the modified growth history.
    
 %===========================FIGURE 5 ==================== 
 \begin{figure}[!t]
  \begin{center}
          \includegraphics[width=3.25in,angle=0]{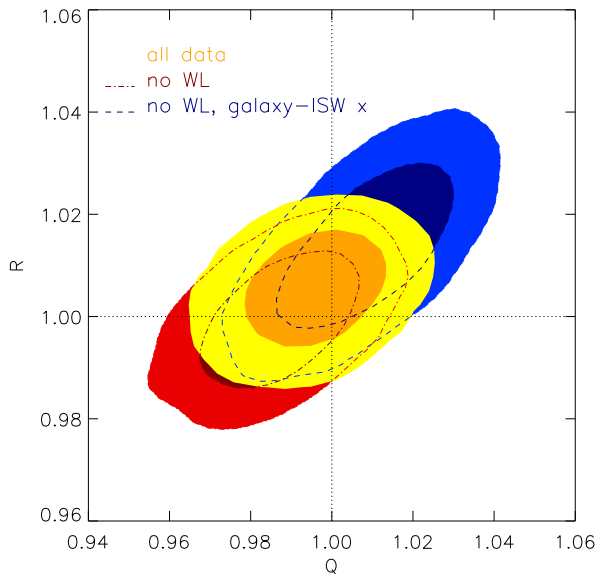}

    \caption{ 68\% and 95\% confidence limits for the time- and scale-independent  modifications in Model I ($Q=Q_0$ and $R=R_0$, $s=0$ and $k_c=\infty$). Constraints are shown for all data combined (forefront, yellow/orange contours), excluding the lensing correlation (red, dot-dashed) and excluding both lensing and galaxy-ISW correlations (blue, dashed).   The effect of a constant modification on correlations around the first acoustic CMB peak leads to tight constraints, limited to deviations $\lesssim 3\%$ from GR at the 95\% confirndence level for all data combined.  Driven by their opposing effects on the well-measured first peak, the principle degeneracy direction is well-described by constant $(Q-R)$.   \label{fig5}}
  \end{center}
 \end{figure}
  %========================================================  

   %===========================FIGURE 6 ==================== 
 \begin{figure}[!t]
  \begin{center}
          \includegraphics[width=3.25in,angle=0]{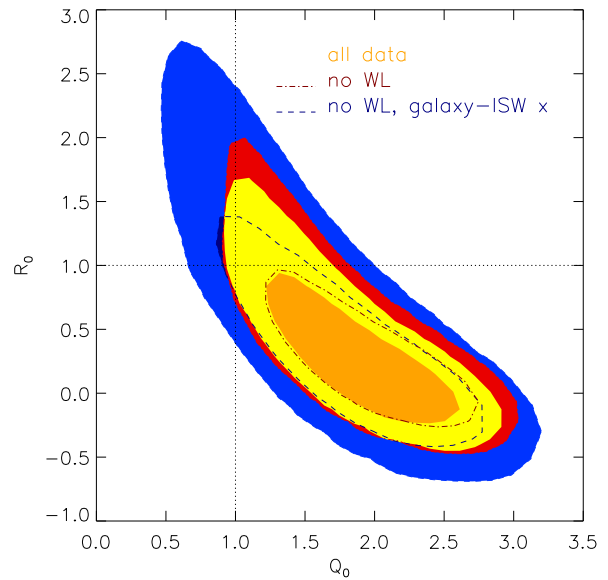}

    \caption{ As in figure \ref{fig5} but giving the 68\% and 95\% confidence limits for allowing scale-independent, but time-evolving modifications in Model I, that become important at late-times ($0\le s\le 3$, $k_c=\infty$).  The principle degeneracy direction is well-described by constant $Q_0(1+R_0)/2$. Constraints on the time-evoving modification are primarily driven by their effect on the cosmic-variance limited, late-time ISW effect in the CMB. As such, they are improved by the inclusion of the galaxy-ISW cross-correlation data, but remain far weaker than for the time-independent modification shown in figure \ref{fig5}. \label{fig6}}
  \end{center}
 \end{figure}
  %========================================================

  We impose flat priors on  six core cosmological parameters: the baryon matter density, $\Omega_bh^2$, the CDM matter density, $\Omega_ch^2$, the angular size of the last scattering horizon, $\theta$, the reionization optical depth, $\tau_{rei}$, the primordial scalar power spectrum index  and amplitude, $n_s$ and $\ln 10^{10}A_s$ respectively. Here $h\equiv H_0/100$ km$s^{-1}$Mpc$^{-1}$ where $H_0$ is Hubble's constant. Following \cite{Lesgourgues:2007te}, we marginalize over absolute calibration uncertainties in the 3 weak lensing bins using 3 nuisance parameters $A,B$ and $C$ as discussed in \ref{data}.  In each scenario, we run 8 independent chains until they satisfy the Gelman-Rubin convergence criteria \cite{Gelman:1992zz}. 

For Model I, we consider both a time-dependent and time-independent scenario. For the time-independent scenario ($s=0$) we assume scale invariance ($k_c=\infty$) and impose flat priors on $0\leq Q_{0,\infty}\leq10$ and $-1\leq R_{0,\infty}\leq10$.  Table \ref{tab1} and figure \ref{fig5} summarize the marginalized 1D and 2D constraints  on $Q$ and $R$ and the principle degeneracy direction.The well measured CMB power spectrum on sub-degree scales provides the dominant constraint, limiting the effect of time-independent modifications to Newton's constant , Q-1$\lesssim3\%$ those from standard gravity at the 95\% confidence level. This is  comparable with constraints on time-independent modifications to the Newton's constant from bounds on the expansion history with the CMB \cite{Robbers:2007ca}. The galaxy-ISW and lensing correlations only provide limited improvements.  As discussed in section \ref{effect}, $R$ and $Q$ have opposing effects on the first peak and the degeneracy direction of the constraints is consistent with this, being well described by $Q_0-R_0$ consistent with zero, and both $Q_0$ and $R_0$ consistent with unity at the 95\% confidence level. 

     %===========================FIGURE 7 ==================== 
 \begin{figure*}[!t]
  \begin{center}
          \includegraphics[width=3.25in,angle=0]{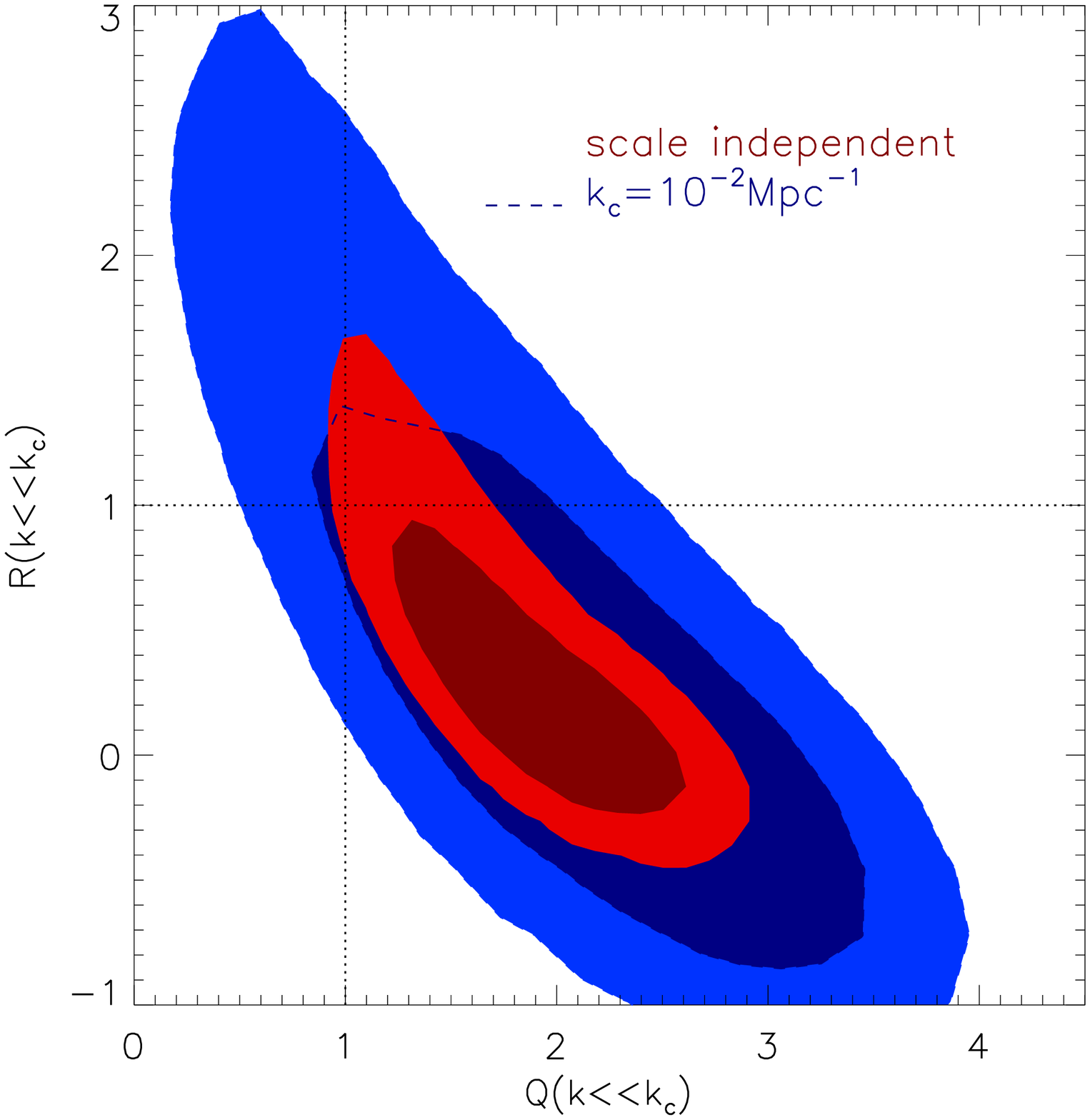}
          \includegraphics[width=3.25in,angle=0]{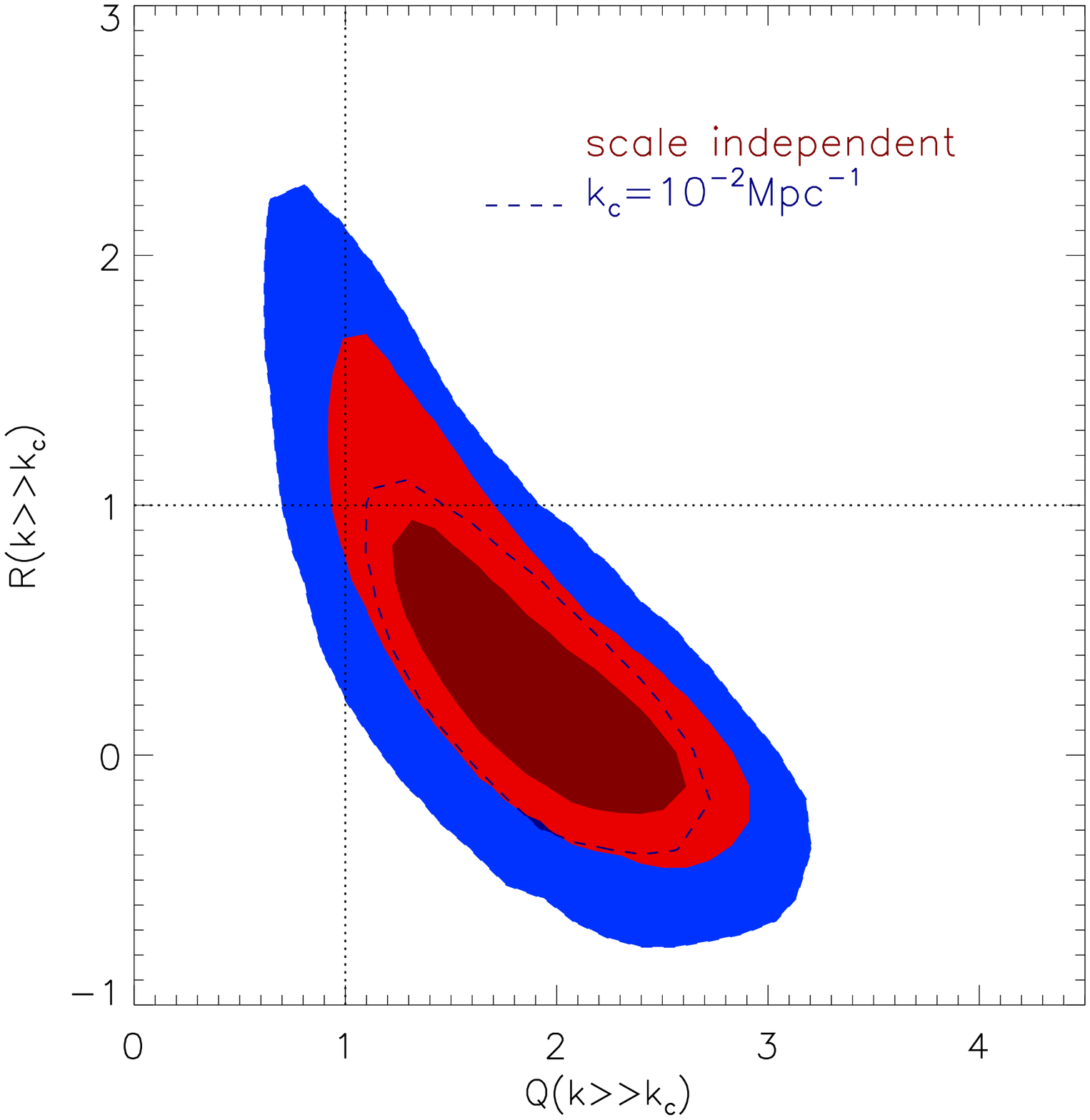}
    \caption{ Impact of allowing different modifications to the growth history  on large and small scales for Model I. The figure shows 68\% and 95\% confidence limits on large [left panel] and small scales [right panel] using all data combined. In each figure, constraints are shown for the scale-independent scenario (forefront, red contours), and the scale-dependent scenario in which distinct deviations from GR are allowed on large and small scales ($Q_0,R_0,Q_\infty$ and $R_{\infty}$ are all varied with cross-over scale $k_c=10^{-2}Mpc^{-1}$ and $0\le s\le3$) (blue, dashed). \label{fig7}}
  \end{center}
 \end{figure*}
  %========================================================

We also investigate time-dependent scenarios for Model I, in which the modification characteristically arises at late times. Here we utilize the degeneracy direction $Q(1+R)/2$ to more efficiently search the parameter space, imposing flat priors on $0\leq Q_{0,\infty}\leq10$  and $0\leq Q_{0,\infty}(1+R_{0,\infty})/2\leq10$, and rejecting combinations with $R<-1$. To ensure the choice of prior did not unduly affect the constraints we also ran MCMC chains using flat priors on $Q_{0,\infty}\sim [0,10]$ and $R_{0,\infty}\sim [-1,10]$ and found no significant change in the results. We consider both scale independent modifications, with $k_c=\infty$ (for which $Q_\infty$ and $R_\infty$ are redundant), and scenarios in which large and small scales are able to vary independently.

Table \ref{tab2} summarizes the 1D marginalized constraints on  late-time evolving modification for which we marginalize over  $0\leq s \leq 3 $. For scale-independent modifications ($k_c=\infty$), one finds that the constraints on the principle measured combination, $Q_0(1+R_0)/2$,  are only marginally affected by omitting the weak lensing data but are more significantly relaxed if the galaxy-ISW correlations are not included in the analysis. This presence of this principle mode is demonstrated by the insensitivity of the constraint to whether $Q_0$ or $R_0$ alone, or both, are allowed to vary, and is clearly shown in Figure \ref{fig6}. 

For scale-dependent modifications  we consider a transition scale of $k_{c}=10^{-2}Mpc^{-1}$. The constraints on the large scale modifications are significantly more relaxed than on the smaller scales, with the dominant constraints coming from the CMB ISW. As shown in Figure \ref{fig7}, the data effectively measures $Q_0(1+R_0)/2$ and $Q_\infty(1+R_\infty)/2$ at large and small scales, respectively.

The change in the maximum likelihoods of the scenarios considered in Model I reflect the degeneracies inherent in the constraints. For all data combined, the fiducial $\Lambda$CDM scenario has best-fit likelihood, ${\mathcal L}$, given by $-2\ln{\cal L}=4070.3$. The scale independent scenario in Model I has a best-fit $-2\ln{\cal L}=4067.7$; though 3 parameters are varied, $Q_0$, $R_0$ and $s$,  only one measured degree of freedom is really introduced, effectively  $Q_0(1+R_0)/2$, while $s$ is unconstrained. The current data does not include sufficient information to constrain any scale dependence in the modification, reflected in no improvement in the best fit in comparison to the scale independent model.

%===========================FIGURE 8 ==================== 
 \begin{figure}[!t]
  \begin{center}
          \includegraphics[width=3.5in,angle=0]{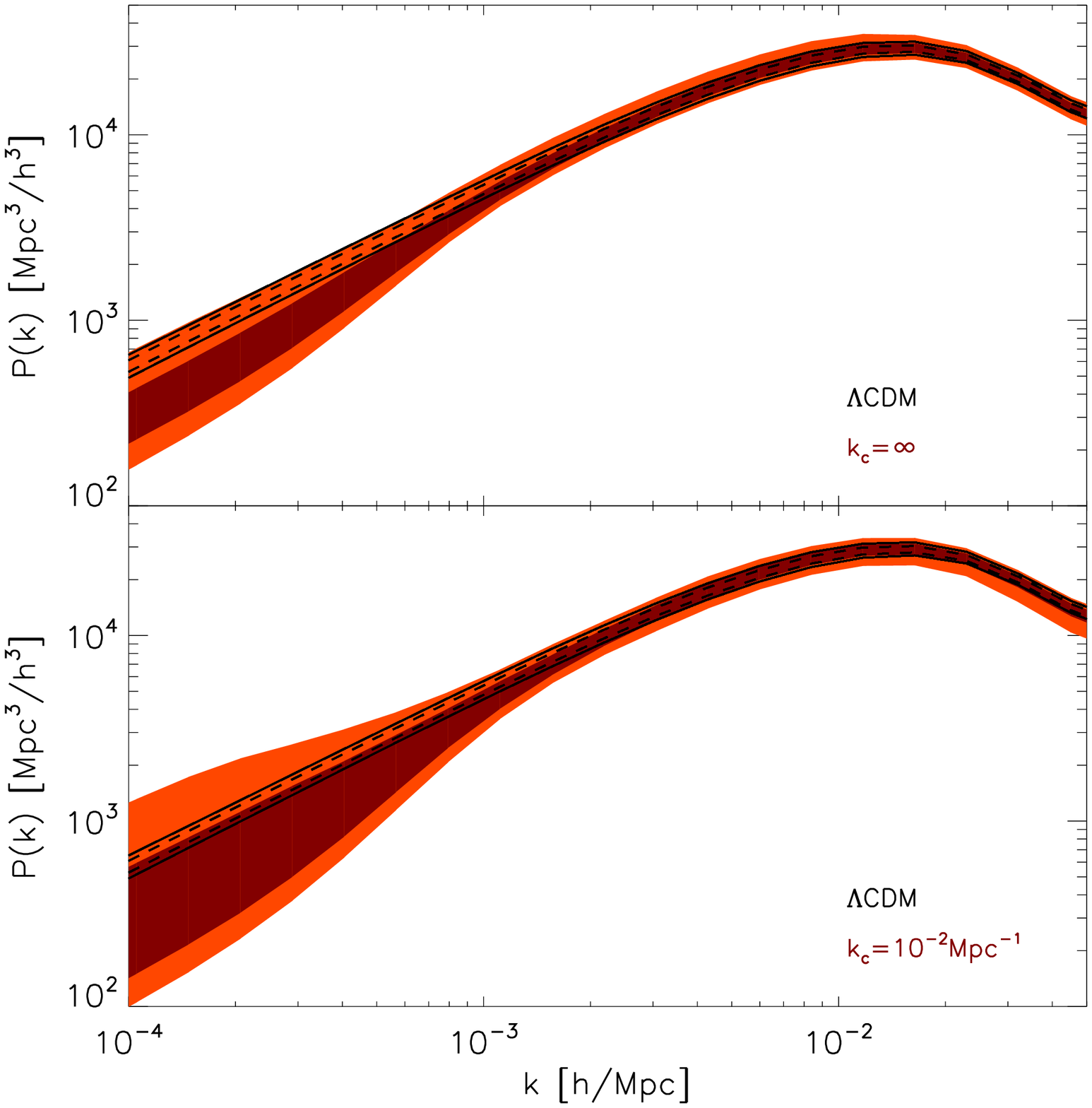}

    \caption{1D marginalized 68\% and 95\% confidence limits on the total matter power spectrum today for $\Lambda$CDM (black contour lines) and the two scenarios using Model I from figure \ref{fig7} (filled red contours). The upper panel shows the constraints for the scale-independent modification while the lower panel shows constraints if both large and small scale growth histories are modified independently. The data allows a broader range of  growth histories, tending to suppress power on large scales relative to the $\Lambda$CDM scenario, constrained by the ISW. By comparison, the matter power spectrum on smaller scales is only marginally effected by the modified growth history. This is even the case when the subhorizon scales are modified independently of the superhorizon scales.   \label{fig8}}
  \end{center}
 \end{figure}
  %========================================================

For the scale-independent and dependent models in figure \ref{fig7}, we calculate the matter power spectrum today in 40 logarithmically spaced bins in $k$ from $10^{-1}-10^{-4} h/Mpc$,  and the growth parameters $f$, $g_0$ and $\gamma$ in 30 bins of width $\Delta z=0.1$ between $z=0$ and 3. These allow us to find the 1D marginalized distributions for the power spectrum and growth parameter values in each bin. 
  
In figure \ref{fig8} we show the 1D marginalized constraints on the matter power spectrum for scale independent and dependent scenarios.  Even in the scenario in which the small scales are allowed to vary separately from the large scales, there is only a small broadening in the amplitude of the spectrum in comparison to that from $\Lambda$CDM. By contrast, the matter power spectrum on large scales can boosted or suppressed relative to $\Lambda$CDM at the 95\% confidence level, with the 68\% confidence levels driven by the low, large scale CMB power. 

In figure \ref{fig9} we show the constraints on the general growth parameters $f$, $g_0$ and $\gamma$  for two comoving scales, $k=0.005/Mpc$ and $0.05/Mpc$. The suppression in large scale growth seen in the power spectrum today is also seen in a lower growth rate (and higher $\gamma$ and $g_0$) at low redshifts. $\gamma$ on large scales can take a dramatically broader range of values than those predicted by $\Lambda$CDM.  Interestingly, on the largest scales the growth rate, $f=d\ln \Delta_c/d \ln a$, can become negative, leaving $\gamma = \ln f/\ln \Omega_m$ undefined. Though $\gamma$ is a good parameterization for indicating a modification to gravity on subhorizon scales, many growth histories consistent with the current data cannot be described by $\gamma$ for scales just below, or at, the horizon scale.

    %===========================FIGURE 9 ==================== 
 \begin{figure*}[!t]
  \begin{center}
          \includegraphics[width=3.5in,angle=0]{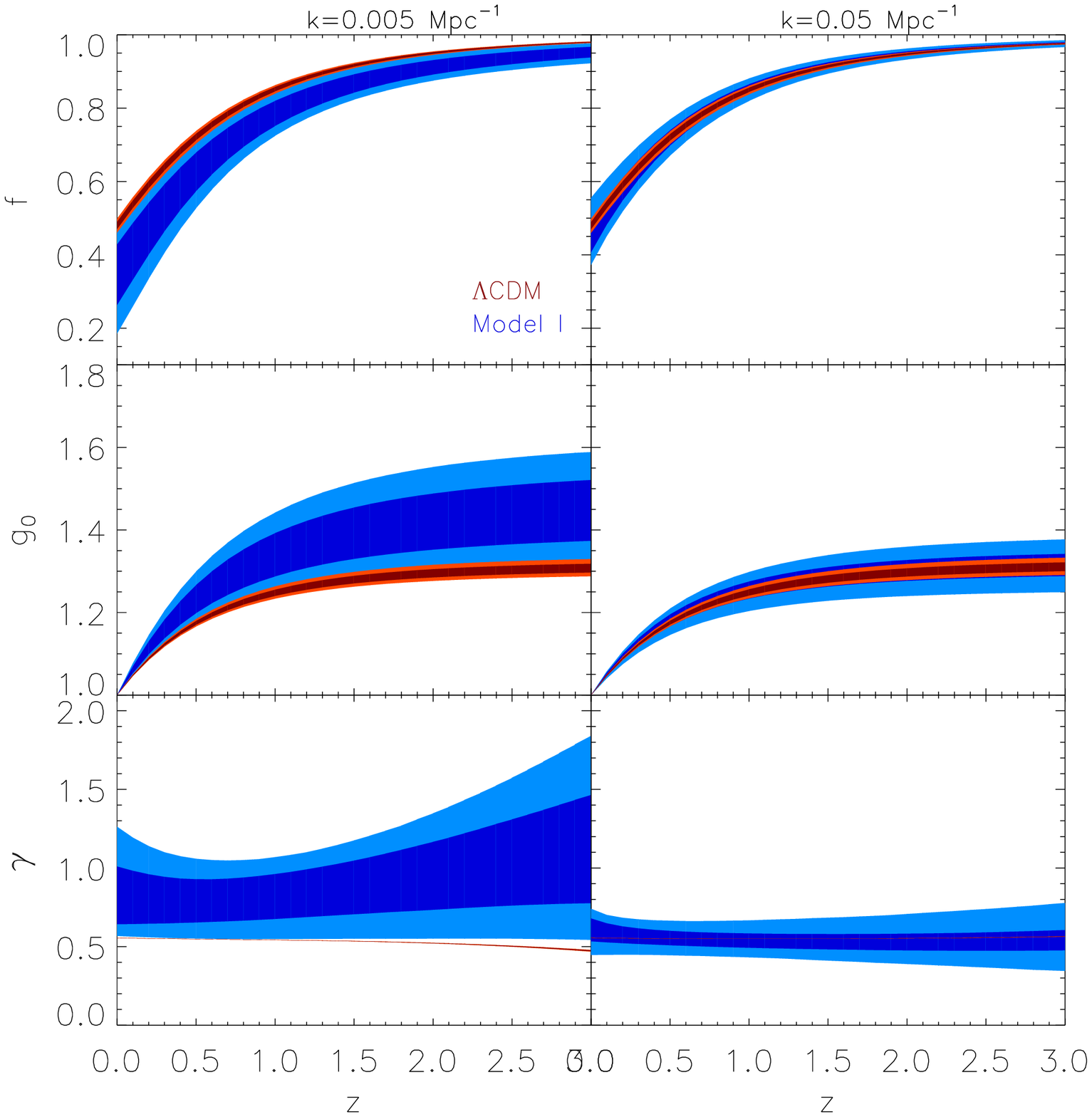}
          \includegraphics[width=3.5in,angle=0]{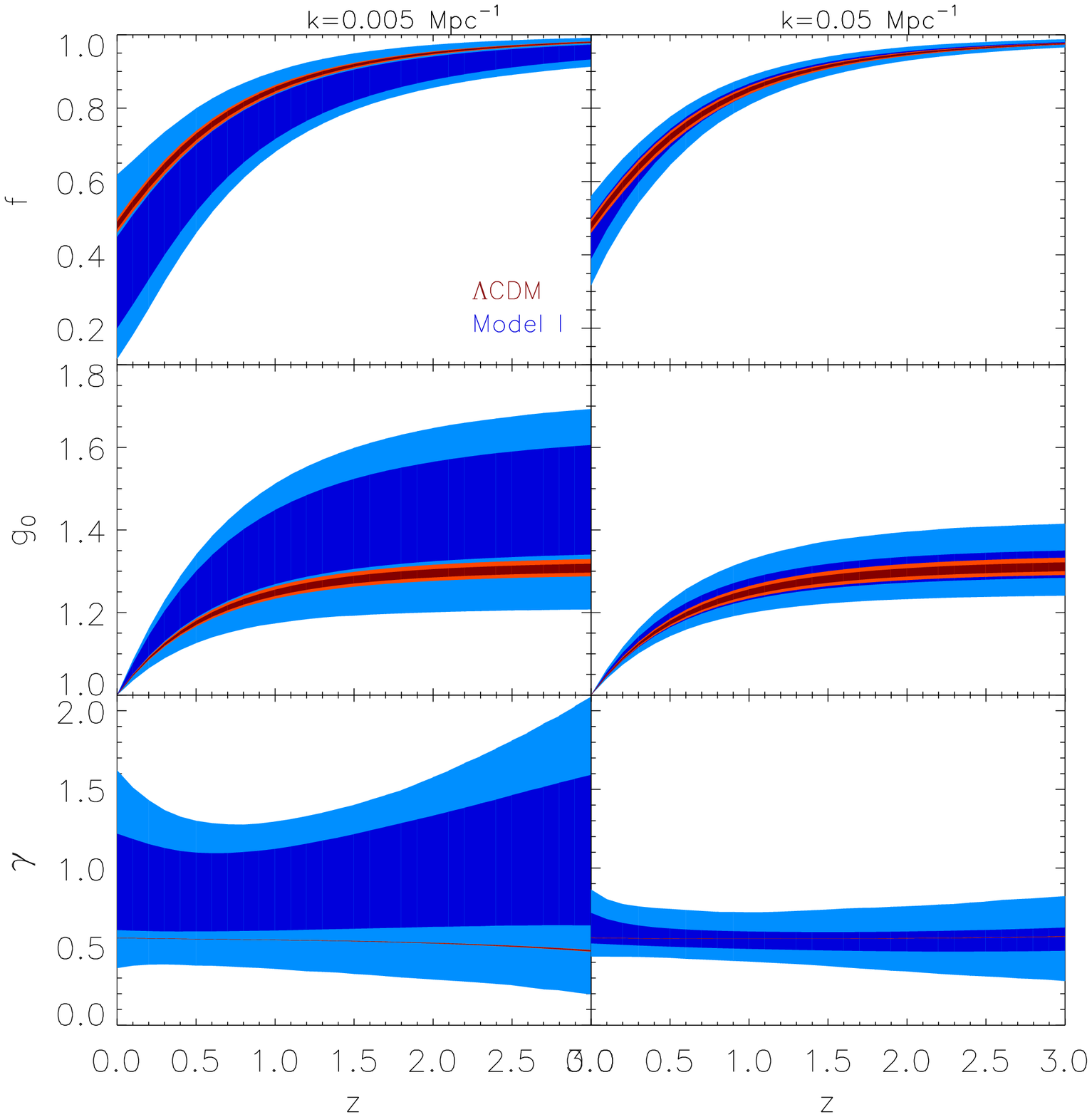}
    \caption{1D marginalized 68\% and 95\% confidence limits for Model I for the growth parameters $f$, $g_0$ and $\gamma$ at two scales $k=5\times10^{-3}Mpc^{-1}$ and $k=0.05Mpc^{-1}$ (left and right panels in each figure respectively). $\Lambda$CDM (red contours) is compared to Model I with $0\le s\le 3$ (blue contours).  [Left figure] Constraints for the scale-independent modification, with $k_c=\infty$. [Right] Constraints for the scale-dependent modification varying $\{Q_0,R_0,Q_\infty,R_{\infty}\}$ with a transition scale $k_c=10^{-2}Mpc^{-2}$. The modified gravity models  can have a markedly broader range of growth rates  than $\Lambda$CDM, especially on large scales. 
     \label{fig9}}
  \end{center}
 \end{figure*}
  %========================================================
   %===========================FIGURE 10 ==================== 
 \begin{figure*}[!t]
  \begin{center}
            \includegraphics[width=3.5in,angle=0]{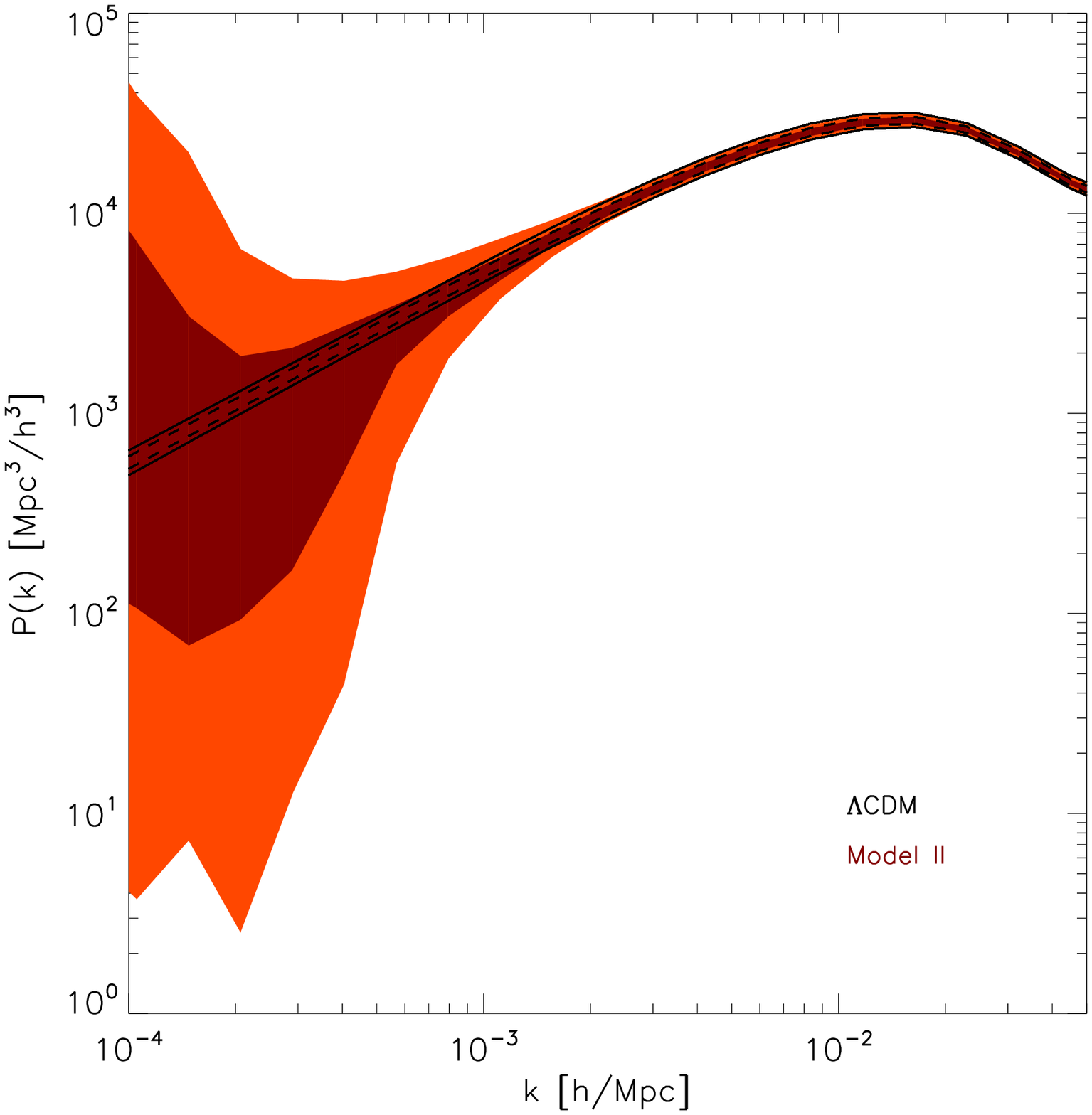}
                    \includegraphics[width=3.5in,angle=0]{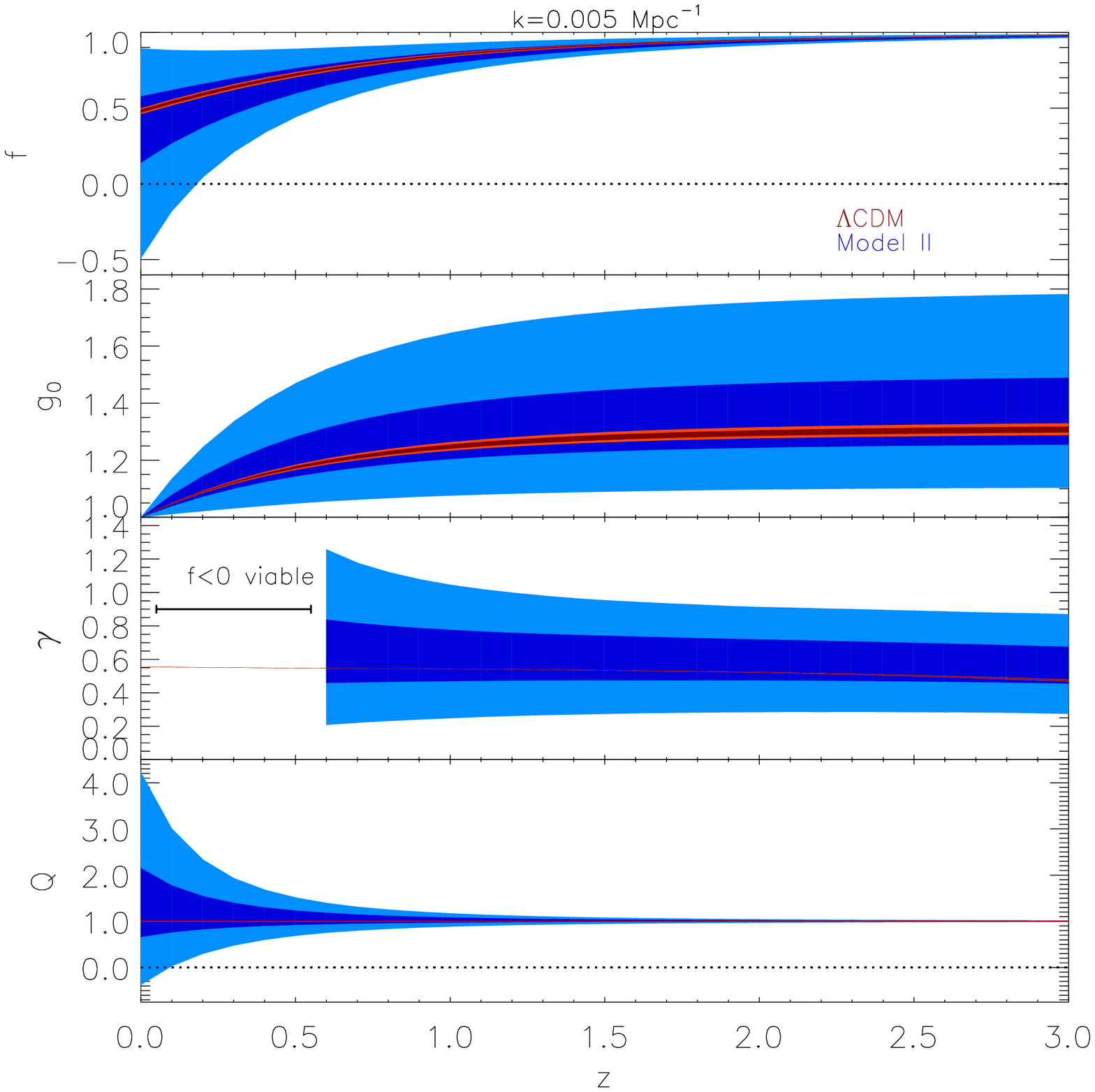}
   \caption{ 1D marginalized 68\% and 95\% confidence limits for Model II  with $k_x=10^{-2}Mpc^{-1}$. [Left] The matter power spectrum for Model II (red contours) compared to $\Lambda$CDM (black full and dashed lines).   [Right]  The growth parameters $f$, $g_0$ and $\gamma$, and $Q$ defined  in (\ref{Qdepert}) at $k=5\times 10^{-3}Mpc^{-1}$ for $\Lambda$CDM (red contours) and Model II  (blue contours). On large scales the suppression of growth can lead to decay of $\Delta_c$, with $f<0$, for which  $\gamma$ is undefined. For redshifts at which the MCMC chains include viable models with $f<0$, we do not give constraints on $\gamma$. \label{fig10}}
  \end{center}
 \end{figure*}
  %========================================================

For Model II, we focus on how modifications effect large scales, considering a cut-off scale of $k_x=10^{-2}Mpc^{-1}$. The modification principally effects modes that have entered the horizon after radiation-matter equality. We impose flat priors on $\Delta_{xH}\sim[-10^{-3},10^{-3}]$, and marginalize over $0\leq s_x\leq 2$ and $0\leq n_x\leq3$.
With all data combined, we find 1D marginalized constraints on $\Delta_{xH}=[-1.5\times 10^{-4}, 3.7\times 10^{-4}]$  at the 95\% c.l. There is not a significant improvement in the maximum likelihood in comparison to $\Lambda$CDM, with $-2\ln{\cal L}$(Model II) =4069.8, reflecting the fact that the constraints are entirely consistent with GR.

In figure \ref{fig10} we show the effect modifications in Model II have on the matter power spectrum and the growth parameters $f, g_0$ and $\gamma$.  The data allows both suppression and growth of large scale power in comparison to $\Lambda$CDM and with a greater amplitude than the parameterization in Model I. Scenarios with a late-time decay of inhomogeneities, with $f<0$ (and undefined $\gamma$) at $k=0.005/Mpc$ are consistent with the data at a 95\% confidence level at low redshifts.

%%%%%%%%%%%%%%%%%%%%%%%%%%%%%%%%%%%%%%%%%%%%%%
\section{Discussion}
\label{discussion}
%%%%%%%%%%%%%%%%%%%%%%%%%%%%%%%%%%%%%%%%%%%%%%

In this paper, we have presented constraints on the growth history of cosmic structure with current cosmological data in light of phenomenological modifications to the Poisson equation and the relationship between the two gravitational potentials, respectively described by two scale- and time-dependent functions, $Q$ and $R$. We consider two parameterizations for these modifications.  The first, Model I, considers monotonically varying $Q$ and $R$ in both time and space. The second, Model II, leaves the relationship between the two potentials unchanged ($R=1$) but allows an additive adjustment to the Poisson equation,  that could Model either a modification to how gravity and matter are related, or equally the presence of an additional inhomogeneous, dark component; from a phenomenological standpoint each interpretation is equivalent, since the effect on structure formation is only detectable indirectly through its impact on the gravitational potential.

We find that time-independent modifications in Model I, through their impact on the well-measured CMB acoustic peaks, are tightly constrained by the data. In terms of the effect on the Poisson equation the constraints translate into the effective Newton's constant being within $3\%$ of its standard value in GR.

An evolving modification that becomes important at late times, as one might expect if it is associated with the onset of cosmic acceleration, is, by contrast, far more weakly constrained; an effective Newton's constant in Poisson's equation today can multiple times its value in GR, rather than a few percent variation, and remain consistent with the data.   The reason for this is that the modifications have their biggest effect on the large scales ($k>k_{eq}$)  that,  at present, principally constrained through the cosmic-variance limited ISW and a small number of data from ISW-galaxy cross-correlations. The inclusion of a wider variety of improved precision, cross-correlations on these larger scales, as will come from upcoming wide-scale surveys,  should greatly improve the constraints on these models.  To precisely compare observations to predictions on these scales will involve considering the effect of gauge issues in relating theory to observables \cite{Yoo:2009au} and the impact of scale-dependence and nonlinearity of the bias on large scales \cite{Dalal:2007cu,Smith:2009pn}, both of which could obscure the effect of an underlying modification to gravity or clustering dark energy.

The modified growth histories considered here can have markedly different behaviors on large and small scales; a boosting of CDM growth on small scales can be tied to a suppression on larger scales. This opens up the question of using sub-horizon and horizon scale observations separately to directly probe for this distinctive signature. 

There is a degeneracy present in constraints from the current data, well described by $Q(1+R)/2$, that prevents one from establishing whether modifications to the growth history arise from $Q\neq 1$ or $R\neq 1$. If we hope to be able to use observational constraints  to  phenomenologically ascertain what is driving any modification to the growth history then breaking this degeneracy is critical. 

The application of tailored estimators built from cross-correlation of galaxy, weak lensing and velocity fluctuations, could directly confront this issue. As such, the combination of deep and wide, imaging and spectroscopic large scale structure surveys, coming online in the next few years, provides the exciting prospect of being able to direct test the cosmic growth history, and the relationship between matter and gravity on cosmic scales.

%%%%%%%%%%%%%%%%%%%%%%%%%%%%%%%%%%%%%%%%%%%%%%
\section*{Acknowledgments}
%%%%%%%%%%%%%%%%%%%%%%%%%%%%%%%%%%%%%%%%%%%%%%
RB would like to thank Niayesh Afshordi, Ghazal Geshnizjani and Jaiyul Yoo for useful discussions. RB and MT's research is supported by NSF CAREER grant AST0844825, NSF grant PHY0555216, NASA Astrophysics Theory Program grant NNX08AH27G and by Research Corporation.  

\appendix

%%%%%%%%%%%%%%%%%%%%%%%%%%%%%%%%%%%%%%%%%%%%%%
\section{Incorporating Modifications to Gravity in the CAMB code}
\label{App-camb}
%%%%%%%%%%%%%%%%%%%%%%%%%%%%%%%%%%%%%%%%%%%%%%

We summarize here how the modified gravity equations discussed in section \ref{deviations} are incorporated into CAMB \cite{Lewis:1999bs}. CAMB uses the synchronous gauge, with metric perturbations $h$ and $\eta$, rather than the  Newtonian gauge.

Using  the notation of Ma and Bertschinger \cite{Ma:1995ey} the synchronous and Newtonian gauge variables are related by
\bea
\phi &=& \eta - \hub \alpha, \ \ \ 
\psi = \dot{\alpha}+\hub\alpha
\eea
where 
\bea
k^2\alpha &=& \frac{\dot{h}}{2}+3\dot{\eta}.
\eea

The synchronous (s) and conformal Newtonian (c) gauge matter perturbations are related by
\bea
\delta_i^{(s)} &=& \delta_i^{(c)}+3\hub(1+w_i)\alpha
\\
\theta_i^{(s)} &=& \theta_i^{(c)}-\alpha k^2
\\
\sigma_i^{(s)} &=& \sigma_i^{(c)}.
\eea

CAMB evolves the metric perturbation $\eta$ and the matter perturbations for each species according to the synchronous gauge fluid equations. To this end, it evaluates the following variables at each time step
\bea
\sigma_{CAMB} &\equiv& k\alpha =\frac{\eta-\phi}{\hub}
\\
z_{CAMB} &\equiv&\frac{\dot{h}}{2k} = \sigma_{CAMB}-3\frac{\dot\eta}{k}.
\eea
where, for the modified gravity model, $\phi$ now comes from the new Poisson equation (\ref{EE000i}), and  the evolution equation for $\eta$ is given by the time derivative of (\ref{EE000i})   coupled with (\ref{EEij}), (\ref{fluid1}) and  (\ref{fluid2}),
\bea
\dot\eta &=& 
 \frac{1}{f_Q} \sum_i4\pi G a^2\rho_i\left[(1+w_i)\left( Q f_{1}\frac{\theta_i^{(s)}}{k^2}+ (Q-1)k^2 \alpha \right)\right.\nonumber 
 \\ && \left.- \left(\dot{Q}+(R-1)Q\hub\right)  \Delta_i\right]\label{etadot}
\eea
where
\bea
f_1 &\equiv & k^2+12\pi Ga^2\sum_i\rho_i(1+w_i)
\\
f_Q &\equiv & k^2+12\pi Ga^2 Q \sum_i\rho_i(1+w_i).
\eea
%%%%%%%%%%%%%%%%%%%%%%%%%%%%%%%%%%%%%%%%%%%%%%
\section{Galaxy auto-correlation analytical marginalization}
\label{App-marg}
%%%%%%%%%%%%%%%%%%%%%%%%%%%%%%%%%%%%%%%%%%%%%%

This section summarizes the analytical marginalization approach for `nuisance' bias and non-linear fitting parameters, $b$ and $Q_g$, used for analyzing the ISW-galaxy cross-correlation data, following the approach of \cite{Ho:2008bz}.

Consider ${\bf d}= \{d^{i}\equiv C_{gg}^{i}(l)\}$,  the vector containing the  galaxy auto-correlation data, with each redshift bin, $i$, and ${\bf C}$ is the associated covariance matrix. To analytically marginalize to find $b_{norm}$ and $Q$ in (\ref{Qbeq}) we construct two theoretical auto-correlation sets
\bea
t^i(l) &=&  \int_0^{\chi_{max}} \frac{d\chi}{\chi^2}W_i(\chi)^2T_{t}(k_l, \chi)^2\Delta_{R}^{2}(k_l)
\\
q^i(l) &=&  \int_0^{\chi_{max}} \frac{d\chi}{\chi^2}W_i(\chi)^2T_{q}(k_l, \chi)^2\Delta_{R}^{2}(k_l)
\\
T_{t}(k, \chi) &\equiv& \tilde{\delta}_{c}(k, \chi) \frac{1}{\sqrt{1+Ak}}
\\
T_{q}(k, \chi) & \equiv& \tilde{\delta}_{c}(k, \chi) \frac{k}{\sqrt{1+Ak}}
\eea
The values of $b_{norm}$ and $Q$ that minimize the $\chi^2$ fit to the data are evaluated using the matrix ${\bf M}$
 and vector ${\bf v}$,
 \bea
{\bf M} &\equiv&\left(
\begin{array}{cc}
\chi^2_{tt} & \chi^2_{tq}
\\
\chi^2_{tq} & \chi^2_{qq}
\end{array}
\right)
, \ \ \ \ \ \
{\bf v} \equiv \left(\begin{array}{c}
\chi^2_{dt} \\ \chi^2_{dq}
\end{array}\right)
\eea
with $\chi^2_{tq} = {\bf t}^T {\bf C}^{-1} {\bf q}$ etc. Then
\bea
\left(\begin{array}{c}
b_{norm}^2 \\
b_{norm}^2Q 
\end{array}
\right)
&=& {\bf M}^{-1}{\bf v}
\eea
with an overall $\chi^2$ fit of
\bea
\chi^2 =\chi^2_{dd} - {\bf v}^T {\bf M}^{-1} {\bf v}.
\eea

%%%%%%%%%%%%%%%%%%%%%%%%%%%%%%%%%%%%%%%%%%%%%%
% BIB
%%%%%%%%%%%%%%%%%%%%%%%%%%%%%%%%%%%%%%%%%%%%%%

\bibliographystyle{apsrev}
\bibliography{ref}

\end{document}